\documentclass{aa}  
\usepackage{graphicx}
\usepackage{txfonts}
\usepackage{lipsum}
\usepackage{subcaption}         
\usepackage{lscape}             
\usepackage{placeins}            
\usepackage{natbib}
\bibpunct{(}{)}{;}{a}{}{,}
\usepackage{color}
\begin{document}
   \title{Distributed accelerators in the jet of Centaurus A: 
  the origin of the spectral hardening of very high energy
  gamma-rays}
   \author{Y. S. Honda\inst{1}
        \and M. Honda\inst{2}
        }
   \institute{Kindai University Technical College, Nabari, Mie 518-0459, Japan\\
             \email{honda@ktc.ac.jp}
             \and Plasma Astrophysics Laboratory, Institute for Global Science, Mie, Japan\\ }
   \date{Received September 6, 2025}
   \abstract
  {We propose the synchrotron self-Compton (SSC) scenario coupled with 
  filamentary jet model, to reproduce the very high energy $\gamma$-ray emissions from Cen A. 
  With reference to self-similarity of knot-like features in the jet, we assume 
  nonuniform magnetic field associated with current filaments having various transverse sizes. 
  For energetic electron production, the diffusive shock acceleration at sites distributed 
  over the kiloparsec-scale jet is considered. We show 
  that maximum Lorentz factor of the electron steadily exceeds $10^{8}$ due to 
  suppression of synchrotron loss of the electrons trapped in weak magnetic field of 
  the thin filaments, and inhomogeneous SSC 
  in the inner jet can dominantly contribute to 
  establishment of the pronounced hardening of $\gamma$-ray flux detected by the H.E.S.S. 
  It is also suggested that the spectral contribution from diffuse regions of 
  the outer jet potentially amounts to the observed Fermi fluxes.}
  \keywords{acceleration of particles --- galaxies: individual (Cen A) --- galaxies: jets --- 
  radiation mechanisms: non-thermal --- turbulence}
  \authorrunning{Honda \& Honda}
  \titlerunning{Gamma-ray spectral hardening of Cen A}
   \maketitle

\section{Introduction}\label{sec:1}
The nearest radio galaxy, Centaurus A [NGC5128, at a distance of 3.8 Mpc   
from us: \cite{2010PASA...27..457H}] 
is known as a powerful $\gamma$-ray emitter, and 
serves for a unique laboratory 
to study the radiative processes in the kpc-scale jets and giant radio lobes. Intriguingly, the High 
Energy Stereoscopic System (H.E.S.S.) detected, in the $\gamma$-ray spectral hump, the unnatural 
hardening above GeV energies \citep{2009ApJ...695L..40A,2020Natur.582..356H}, which exceeds the flux level extrapolated from the 
early Fermi-LAT data in the sub-GeV range \citep{2010ApJ...719.1433A}. The H.E.S.S. suggests the Lorentz factor of
accelerated electrons reaches the order of $10^{8}$. 
Although the single-zone SSC scenario \citep{2001MNRAS.324L..33C} can adequately reproduce the spectra of 
Cen A up to sub-GeV range detected by the EGRET on CGRO \citep{1998A&A...330...97S}, it is not accountable for 
the very high energy (VHE; 100 GeV-100 TeV) $\gamma$-ray fluxes detected by the H.E.S.S. \citep{2020Natur.582..356H}. 
\citet{2010ApJ...719.1433A} constructed a multi-band spectrum of Cen A core below $\sim$10 GeV using a single-zone 
SSC model, but required an alternative parameter set so as to explain the early H.E.S.S. data beyond 100 GeV 
\citep{2009ApJ...695L..40A}. \citet{2008A&A...478..111L} proposed a model of multiple blobs having similar size, positioned 
at various angles with respect to the jet axis and moving in different directions to the line of sight.   
Their model in which the electron acceleration site was supposed to be in the jet-forming region
could also explain the H.E.S.S. spectrum, but the supposition is somewhat incompatible with the latest 
observational result that the $\gamma$-rays are produced in the whole jet \citep{2020Natur.582..356H}.
Assuming more intense magnetic field strength than in the conventional SSC modeling, 
\citet{2014A&A...562A..12P} considered the VHE emission that stems from hadronic component, on the basis of the  
single-zone SSC model. 
Inverse Comptonization (IC) of dust emissions by electrons accelerated at velocity shear \citep{2017ApJ...842...39L,
2025PhRvD.111b3024M} might be a possible process of the VHE $\gamma$-ray production, as compatible 
with the H.E.S.S. result \citep{2020Natur.582..356H}. In any case, the substantial observations prompt us to elucidate 
fundamental mechanism of particle acceleration responsible for the low energy spectral hump from radio to 
X-rays being arguably reflected in the $\gamma$-ray hump. 

The broad $\gamma$-ray image is reminiscent of ubiquity of 
X-ray knots in the jet. The high-resolution X-ray images \citep{2002ApJ...569...54K} reveal that 
the X-ray knots are embedded in radio knots \citep[][hereafter BFS83]{1983ApJ...273..128B} and similar 
in form to them, implying the common acceleration mechanism for electrons emitting the synchrotron photons.
In this aspect, we conceive that the X-ray knots, 
which can be associated with shock waves \citep{2002ApJ...569...54K}, are  
major origin of the VHE $\gamma$-ray fluxes. The shock is generated by 
collision of flowing plasmas, and could realize the 
diffusive shock acceleration (DSA) of electrons and positrons {\it in situ} 
\citep{1983RPPh...46..973D,1987ApJ...322..643B}. 
On the other hand, the violent situation is known to be unstable for the 
electromagnetic current filamentation (Weibel-like) instability 
\citep[e.g.,][]{2004PhRvE..69a6401H,2007ApJ...668..974K}. 
Nonlinear development of the turbulent state results in the nonuniformity 
of magnetic fields \citep{2003ApJ...596L.121S}, which should be incorporated 
in modeling of the accelerator as the synchrotron emitter.  

This motivates us to consider an inhomgeneous SSC process in the kpc-scale jet as a possible mechanism 
of the VHE $\gamma$-ray production. For the scale-invariant magnetized current filaments 
reflecting self-similarity, the magnetic field strength depends on the transverse size 
$\lambda$ of the filaments, conforming to the turbulence spectrum of $B_{k}^{2}/8\pi$ with $k$ 
being $2\pi/\lambda$ \citep[e.g.,][]{1979PhFl...22..866M}. The field strength of 
the thinner filament is weaker than that of the largest filament comparable to the global 
mean inferred from energy equipartition with photon fields. Then, we can expect, in accord with the 
foregoing X-ray observation, that electrons in the thinner filaments be accelerated 
to higher energies, owing to the suppression of synchrotoron cooling.   
In simple modeling with evenly fragmented multi-blobs, it is hard to postulate such 
weak magnetic field without violating consistency with the global equipartition. 
Concerning the comparison with the spectral measurement, the point is that the brighter knots/subknots close to 
the nucleus are obscured by thick dust lane in the infrared to ultraviolet band. This makes the direct comparison 
with their intrinsic non-thermal spectra 
difficult. Thus, building the SSC model accountable for the VHE fluxes is directly linked to the key problem: how 
to reasonably construct, in the inner jet, the synchrotron X-ray spectrum including the extended high energy 
tail and its cutoff, as reconciled with the X-ray fluxes measured in the Chandra band \citep{2002ApJ...569...54K}.

In this paper, we adopt a phenomenological approach to construct the radiation spectrum 
in the radio to $\gamma$-ray range. Rather than solving the kinetic equation with a monoenergetic injection term 
to obtain temporal evolution of the electron energy distribution, we focus on the detailed 
internal structure of the jet, considering the steady distribution. We address 
that the inhomogeneous SSC processes in the filamentary jet-knots 
can be a favored candidate for major mechanism of the  
VHE $\gamma$-ray production.
The model is based on the DSA of electrons trapped in the magnetic field of the scale-invariant filaments, 
and the competition with energy loss processes is assessed quantitatively for the filaments each. It is 
demonstrated that in the thinner filaments with the weaker magnetic fields, the electrons are accelerated 
to higher energy, and in the bright X-ray knots of the inner jet, the maximum Lorentz factor reaches $10^{9}$ 
in a critical region where synchrotron cooling time of the electrons is comparable to timescale of their 
local escape from there. 
In the DSA sites distributed in the kpc-scale jet each, the theoretical 
synchrotron spectra are constructed by numerically superposing the 
fluxes from the numerous filaments. Taking account of radiative cooling 
of the filaments and the SSC process in the knots A and B, 
we reproduce the $\gamma$-ray hardening observed by the H.E.S.S. The 
physical parameters such as the compression ratio of scattering center, 
magnetic field strength, and so on, 
required for spectral reproduction are fixed within a narrow range. 
If we consider the contribution from diffuse regions of outer jet far away from central engine,    
the data observed by the Fermi-LAT, as well, can be almost reproduced above GeV.

In the subsequent Sect. \ref{sec:2}, we provide an explanation of the filamentary jet model,  
in terms of its application to the knots of Cen A; that is, the self-similar structure formed by 
the generation and nonlinear development of the current filaments (Sect. \ref{ssec:21}), 
the DSA and energy loss processes in the filaments (Sect. \ref{ssec:22}), 
semi-numerical method of constructing the synchrotron spectrum (Sect. \ref{ssec:23}), 
and conceivable IC processes (Sect. \ref{ssec:24}). In Sect. \ref{sec:3}, the reproduced 
spectra of knots and a diffuse region are shown, and compared with the H.E.S.S. and Fermi fluxes. 
Section \ref{sec:4} is devoted to conclusion and discussion on spectral variability.

\section{Theoretical modeling}
\label{sec:2}
\subsection{Filamentary knots and the self-similarity}
\label{ssec:21}
\begin{figure}[t]
\begin{center}
\includegraphics[trim  = 0 0 0 0, width=8cm,clip]{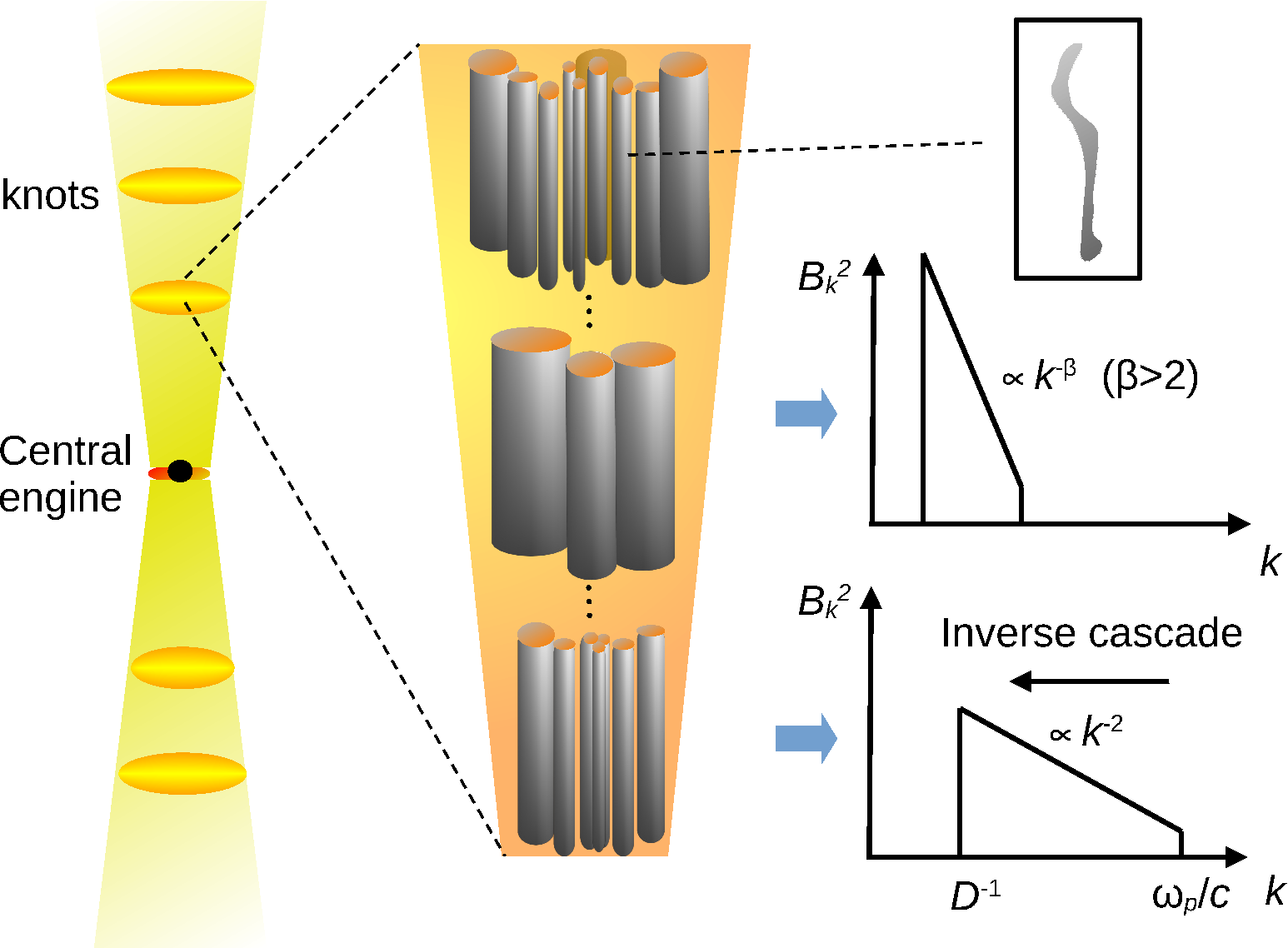}    
\end{center}
    \caption{The schematic view of the current filamentation and coalescence at a knotty region. When  
    plasmoids ejected intermittently from the central engine catch up with the preceding flow, 
    the bulk breaks up into numerous separate fragments. The filaments are represented by 
    many pipes; the inset illustrates a sample of the realistic one bent or twisted. The filaments soon 
    coalesce due to the attractive force 
    acting on the currents in the same direction. These processes actively repeat, because of multiple shocks. 
    The corresponding spectra of turbulent magnetic fields are depicted on the right side. The 
    filamentation is driven at the scale of plasma skin depth ($c/\omega_{p}$), and the magnetic field 
    energy accumulates at the largest scale ($D$), via the inverse cascade in the $k$-space 
    (leftward arrow).}
    \label{fig:processes}
\end{figure}

The jet bulk may be composed of electron-positron plasma \citep[e.g.,][]{2000ApJ...545..100H},  
but it is enough for our purpose to consider the acceleration of (emissions from) electrons 
only, relying on the charge symmetry. 
The filamentation instability is expected in the jet-knots/blobs \citep{2007ApJ...654..885H}, 
and in huge-current carrying jet as well \citep{2002ApJ...569L..39H}; the former involving shocks is
in on the present issue. In our view, each of the radio knots (such as A, B, $\cdots$) 
having the jet width  
contains multiple shocks, as intermittent activities of central engine lead to successive 
impingement of ejected plasmoids on the preceding slower ones.  
We envisage that inside the knot, the filamentation driven by the shocks, coalescence of the current filaments, and 
condensation of magnetic energy \citep{2000Phpl....7.1302H,2003ApJ...596L.121S} are taking place
everywhere, so that the filaments are present not only behind the shocks, but also ahead.  
The power-law spectrum of the turbulent magnetic fields $B_{k}^{2}\propto k^{-\beta}$ 
effectively appears in the inertial range resulting from the inverse cascade. That is, due to the 
attractive force acting on the currents in the same direction, some of them coalesce each other into the larger 
one. The expected self-similar structure is in accord with the observed features of subknots-in-knot 
\citep{2002ApJ...569...54K} and subsubknots-in-subknot \citep{2003ApJ...593..169H,2019ApJ...871..248S}.
Figure \ref{fig:processes} schematically shows the dynamical picture of the filamentary knot
as a filaments ``woods''.
When the filamentation is actively 
occurring, the turbulent spectral index reflects an ideal state of two-dimensional turbulence 
of transverse magnetic fields, to be $\beta=2$ \citep{1979PhFl...22..866M}. In the advanced stage of turbulence, 
the value of $\beta$ likely increases \citep{2003ApJ...596L.121S}, but the subsequent 
filamentation will reset it to 2. 
Because of the multiplicity of the shocks, these processes repeat, to virtually equilibrate inside the knot.
In knots of the outer jet, weaker shocks and/or longer interval among the shocks 
may bring about larger effective $\beta$-value, if at all.

After transit of a shock train, i.e., in post-knot region, the filamentation is no longer 
active, so that the accumulation of larger filaments be prominent. The phase of energy condensation 
at the largest scale of the jet wide could be also expected  
in the diffuse regions of the outer jet. The situation in which small filaments are swept away, 
leaving only a few large filaments, is similar to the special case in which is 
solely a filament, as compared to the single-zone model. 

Meanwhile, the plasma will evolve toward an energy-relaxed state magnetohydrodynamically, 
to self-organize the force-free helical flow \citep{1957ApJ...126..457C,2009ApJ...703L.104B}, 
which is impinged upon by the leading edge of subsequent shocks.  
The theory suggests that in lower luminosity region between knots, the jet may lose 
its sharp shear, having rather smooth velocity distribution. 
The specific configuration likely degrades efficiency of the shear acceleration, but acts as a 
pre-stage accelerator to inject seed electrons into the forthcoming shocks.

\subsection{Diffusive shock accelerator of electrons}
\label{ssec:22}
Locally ordered magnetic fields of the filaments are inevitably disturbed by 
the Alfv\'{e}n waves. We assume that the energy density ratio (denoted as $b$) of  
the fluctuating magnetic fields to the mean field $B(\lambda;\beta)$, is smaller
than unity, and constant independent of $\lambda$. The 
gyrating electrons bound to the mean field of filament are resonantly scattered when the condition 
$r_{g}\sim k'^{-1}$ is satisfied \citep{1983RPPh...46..973D}, where $r_{g}(\gamma_{e},B)$ 
and $k'(\geq k)$ are the gyro-radius of the electron with the Lorentz factor $\gamma_{e}$ 
and wave number of the Alfv\'{e}nic fluctuations, respectively. The successive small-angle 
deflections enable the electrons to go back and forth across the shock statistically; 
this is a subject of the Fermi-based DSA mechanism. In the plasma configuration as argued in 
Sect. \ref{ssec:21}, the magnetic field lines tend to be largely inclined, 
rather than parallel, with respect to the shock normal direction. 
Besides, the inclination angle is distributed when the filaments are wiggling (cf. Fig. \ref{fig:processes} 
inset) as seen in observations \citep{2018NatAs...2..472G,2023NatAs...7.1359F,2024A&A...692A..48R}. In this aspect, 
an average of the inclination angle 
upstream the shock is considered, and its value is set 
to the de Hoffmann-Teller limit: $\theta=\cos^{-1}(U/c)$ \citep[with the shock speed of $U=0.5c$ 
comparable to the speed of proper motion: ][]{2003ApJ...593..169H}, in which 
the acceleration efficiency is maximum \citep{2005MNRAS.362..833H}. 
Taking all this into account, the characteristic acceleration time of the electron is given 
by $t_{\rm acc}=[3\eta r/(r-1)](r_{g}/c)$, where $r$ is the scattering center compression 
ratio, which is precisely related to the gas compression ratio \citep{Schlickeiser02}, and
\begin{equation}
\eta=\frac{3}{2b}\left(\frac{\lambda}{2r_{g}}\right)^{\frac{2}{3}}
\left[1+\frac{r}{(\cos^{2}\theta +r^{2}\sin^{2}\theta)^{3/2}}\right],
\label{eqn:eta}
\end{equation}
which is valid when $k'$ is of the inertial range of the Kolmogorov spectrum 
\citep{1987ApJ...322..643B}.

As concerns the processes that limit the acceleration, it is crucial for the temporal 
limit to take synchrotron radiation loss into account, providing that energy density of the 
magnetic field $u_{\rm m}=B^{2}/8\pi$ is locally comparable to that of synchrotron 
photons $u_{\rm ph}$, or more.
The extended argument on photon fields is given later in Sect. \ref{ssec:24}. 
We take no account of adiabatic expansion effect 
of the acceleration regions, since the magnetic
self-collimation \citep{2002ApJ...569L..39H} will significantly prevent
them from expanding at least radially outward. Then, for the electrons bound to stronger 
magnetic fields of larger filaments, 
the acceleration is limited by the synchrotron cooling time of 
$t_{\rm syn}=3m_{e}c/4\sigma_{\rm T}u_{\rm m}\gamma_{e}$, where 
$\sigma_{\rm T}$ is the cross section of Thomson scattering, and the other notations are standard.
The equation of $t_{\rm acc}=t_{\rm syn}$ provides the solution 
for $\gamma_{e}$, namely, the maximum Lorentz factor (denoted as $\gamma^{*}$) 
determined by the temporal limit. On the contrary, for the electrons in  
smaller filaments, the escape loss overcomes the synchrotron loss; note here 
that the "escape" does {\it not} mean running away from the jet wide. In other words, 
the spatial limit $r_{g}=\lambda/2$ determines $\gamma^{*}$ in that regime. 
It is remarked that the local escape could successively occur until the electrons are 
trapped in a filament larger than $\lambda_{\rm c}$, to be further energized or 
radiatively cooled. Anyhow, there exists the critical size of filament $\lambda_{\rm c}$ 
at which the energy restriction of electrons switches from the temporal to the spatial limit 
\citep{2007ApJ...654..885H}. 
The achievable maximum of $\gamma^{*}$ is realized 
at $\lambda_{\rm c}$, to give the synchrotoron cutoff frequency.  
We numerically evaluate $\gamma^{*}$ for various filament sizes $\lambda$,
whereupon construct the synchrotron spectrum by the method explained below.

\subsection{Semi-numerical method of constructing theoretical synchrotron spectra}
\label{ssec:23}
\begin{figure}[t]
\begin{center}
\includegraphics[trim=0 0 0 0, width=8cm,clip]{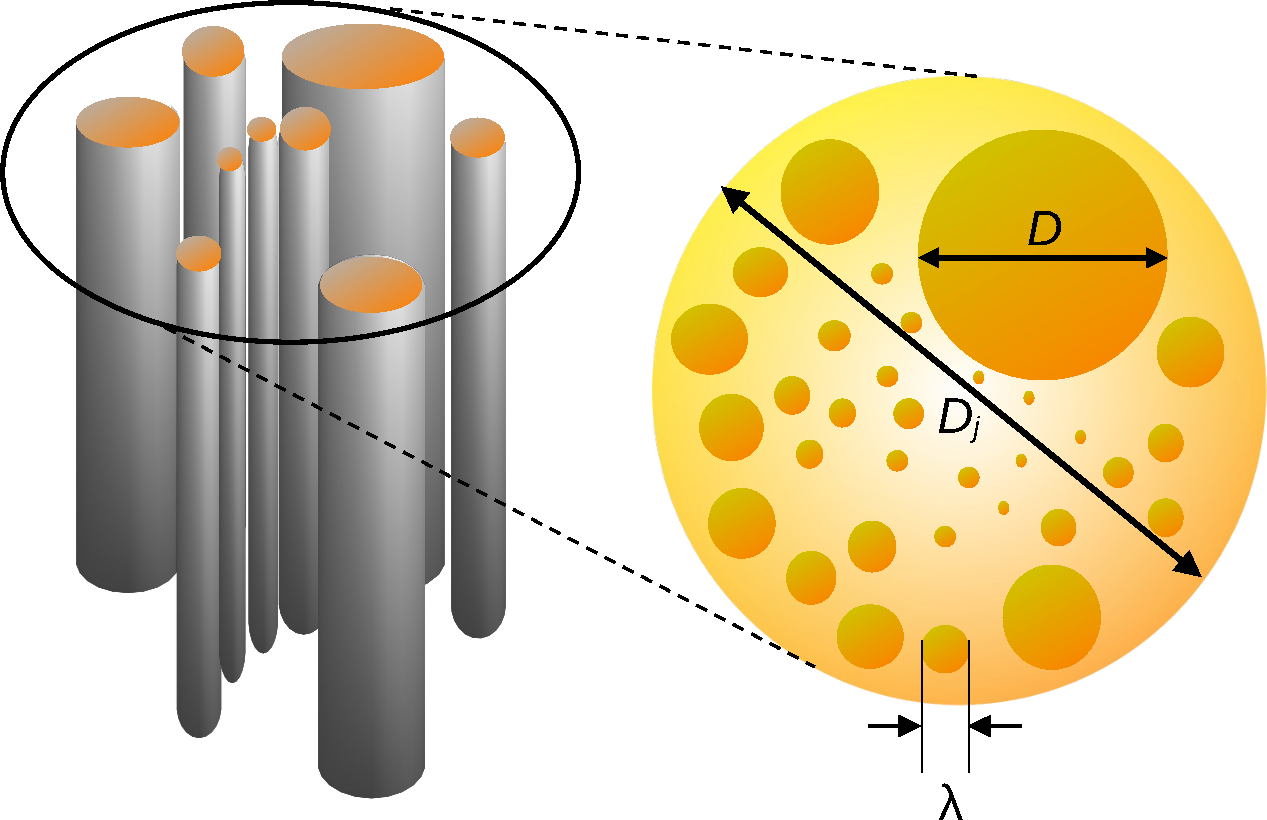}  
    \vspace{1em}
\end{center}
\caption{A transverse cut of the knotty region that contains, with a filling rate less than unity, 
    filaments having various transverse sizes $\lambda$. Note the relation of $\lambda\leq D<D_{j}$, 
    where $D$ is diameter of the largest filament and $D_{j}$ corresponds to diameter of the jet.}
\label{fig:cutting-edge}
\end{figure}

Theoretical maximum value of $\lambda$ is limited by the observed transverse size of the 
knots/diffuse region, both of which are denoted as $D_{j}$. Here, the maximum of $\lambda$, 
denoted as $D$, is given by $f_{\rm cr}D_{j}$, where $f_{\rm cr}$ is a 
parameter less than unity, which stands for correlation reduction of the filamentary 
turbulence.
On the other hand, theoretical minimum of $\lambda$ is limited by the 
plasma skin depth of $c/\omega_{p}\sim 5\times 10^{5}n_{e}^{-1/2}$ cm (cf. Fig. \ref{fig:processes}), 
where $\omega_{p}$ and $n_{e}$ 
(in cm$^{-3}$) are the angular frequency of plasma oscillation 
and number density of electrons (typically, $n_{e}\sim 10^{-2}$ cm$^{-3}$; BFS83), 
respectively. Between $D$ and $c/\omega_{p}$, the scale of $\lambda$ is supposed 
over many orders of magnitude, as compared to the powers of tenth structure of radio jets 
\citep{1984ARA&A..22..319B}. The related self-similar feature is illustrated in Fig. \ref{fig:cutting-edge}, 
along with the relation between $D$ and $D_{j}$. 
In the modeling, introducing the effective minimum $d(\geq c/\omega_{p})$, we have $D/d=10^{n}$
with $n$ being allowable up to $\sim 15$.

The contribution of various filaments to the flux is considered as follows.
In the $\lambda$-space, numerical filaments with the total number $N$ are distributed 
at equal interval $\Delta=n/N$ on an exponent, having no characteristic length-scale. 
Let $S_{i}$ be the area of cutting edge of the $i$-th filament and $\mathcal{F}(S_{i})$ be 
the size distribution of the filaments.
As the specific form of $\mathcal{F}$ is unknown yet, we assume that it has a form such as
$\mathcal{F}(S_{i})=(S_{i}/S_{0})^{-q}$, where $q$ is a constant satisfying $0\leq q<2$.
Then, we {have} the following requirement:
\begin{equation}
\sum_{i=0}^{N}\mathcal{F}(S_{i})S_{i}=\frac{S_{0}\{1-[10^{-(2-q)\Delta}]^{N+1}\}}{1-10^{-(2-q)\Delta}}
=\frac{\pi}{4}f_{p}D_{j}^{2},
\label{eqn:sum}
\end{equation}
\noindent
where
$S_{0}=(\pi/4)D^{2}$
with $D=f_{\rm cr}D_{j}$ is
the area of the largest filament and $f_{p}$ is the
filling rate of filaments in the
jet, which satisfies $f_{\rm cr}^{2}<f_{p}<1$.
From Eq. (\ref{eqn:sum}) for $N\gg 1$, we have
$f_{\rm cr}^{2}/f_{p}=1-10^{-(2-q)\Delta}$, 
which is written as $(2-q)n=-N\log_{10}(1-f_{\rm cr}^2/f_{p})>0$.
In calculation of the DSA and construction of synchrotron spectrum,
the parameter $f_{p}$ is merely involved in the logarithmic argument,
whereas varying the value of $f_{\rm cr}$ directly affects shape of the spectrum
[the break frequency appears proportional to $f_{\rm cr}^{-1}$; cf. Eq.~(\ref{eqn:b})].
For simplicity, therefore, we set to as $f_{p}=0.5$.
Then, regarding the unknown quantity of $(2-q)n$ as invariant of the turbulent dynamics,
we change $N$ when varying the $f_{\rm cr}$-value in spectral fitting.
\begin{figure}[t]
\begin{center}
\includegraphics[trim= 0 0 0 0, width=8cm,clip]{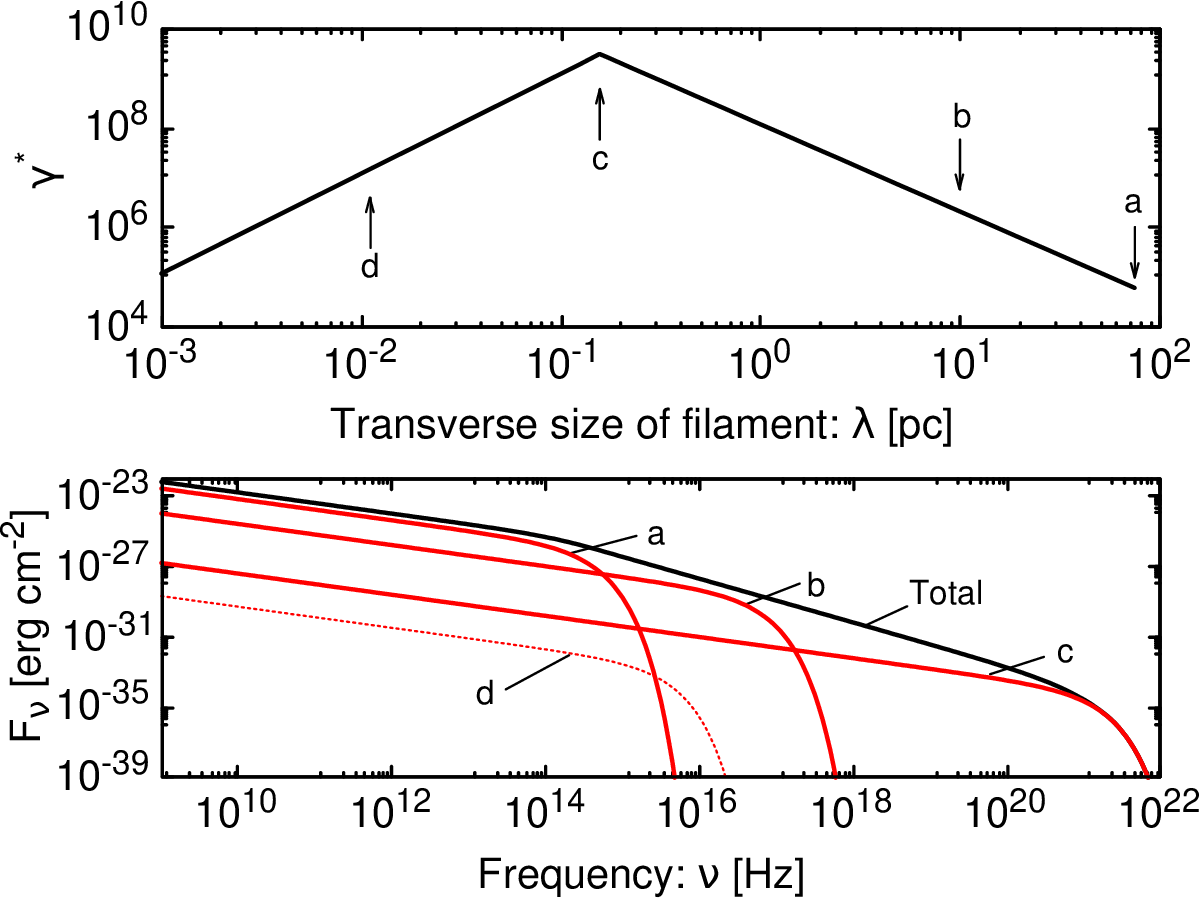} 
\end{center}
\caption{The maximum Lorentz factor of electrons accelerated in the filaments having  
    various sizes $\lambda$ in a knot (top) and flux density due to contribution from 
    sampled filaments (the labels 'a-d' correspond; bottom). 
    The solid black curve (labeled as 'Total') shows the flux that is the sum of contributions 
    from all filaments in the considered $\lambda$-range. For an explanation, see the text.}
\label{fig:gamma-flux}
\end{figure}

As for the electron acceleration, the DSA mechanism generates the energy distribution 
of electron density $n(E)dE\propto E^{-p}dE$ with $p=(r+2)/(r-1)$, where
$p$ is assumed to be constant in the emission regions each. Note here that 
$p$ is not decomposed into 
multi-values; the steady electron distribution is presumed with continuous 
injection, instead of solving the kinetic evolution. We consider synchrotron emissions from the electrons in the 
$i$-th numerical filament having the size $\lambda_{i}$, and then, sum them all.
The relativistic beaming effect is ignored \citep{2020Natur.582..356H}.
Provided $n(E)=c_{i}E^{-p}$ with $c_{i}$ being a constant for the $i$-th filament,
the number density of electrons injected into the $i$-th is
given by $n_{i}\simeq [c_{i}/(p-1)]E_{{\rm min},i}^{-(p-1)}$,
where $E_{{\rm min},i}$ is the energy of the injected electrons.
The emissivity per unit volume of the $i$-th can be approximately expressed as
$J_{i}(\nu)\approx c_{i}B(\lambda_{i})^{\frac{p+1}{2}}\nu^{-\frac{p-1}{2}}
\exp(-\nu/\nu_{i})$ \citep[e.g.,][]{Longair94}, where
$B(\lambda_{i})=B_{\rm m}(\lambda_{i}/D)^{(\beta-1)/2}$ with $B_{\rm m}=B(\lambda_{0}=D)$,
and $\nu_{i}=(3/4\pi)e\gamma_{i}^{2}B(\lambda_{i})/m_{\rm e}c$ with 
$\gamma_{i}=\gamma^{*}(\lambda_{i})$.
Here, a factor of order unity (dependent on $p$ etc.) has been ignored.
To evaluate the luminosity $L_{\nu,i}$ of the $i$-th,
the $J_{i}(\nu)$ must be multiplied by the area of the filament: $S_{i}$ and
the number of filaments of that size: $\mathcal{F}(S_{i})=\mathcal{F}_{i}$.
That is, we have
$L_{\nu,i}\approx\kappa_{i}B(\lambda_{i})^{\frac{p+1}{2}}\nu^{-\frac{p-1}{2}}\exp(-\nu/\nu_{i})$,
where the $i$-dependent coefficient is explicitly given by $\kappa_{i}=c_{i}\mathcal{F}_{i}S_{i}=n_{i} 
(p-1)E_{{\rm min},i}^{(p-1)}\mathcal{F}_{i}S_{i}$.
From this follows: for the likely $p\simeq 2$ that corresponds to $r\simeq 4$, we have
$\kappa_{i}\simeq\mathcal{F}_{i}S_{i}n_{i}E_{{\rm min},i}$,
and as is, $\kappa_{i}$ can be claimed to almost stand for 
the total energy of electrons injected into the filaments having the $i$-th size.
In any case, the sum $\sum_{i=0}^{N}L_{\nu,i}$ provides, in effect, the quantity
as could be compared with observed flux density $F_{\nu}$ of the emission region.
When decently assuming $\kappa_{i}\sim const.=\kappa$,
the summation is significantly simplified; e.g., both the cases of
$\kappa=S_{i}n_{i}E_{{\rm min},i}$ for given $(q,n)=(0,\mathfrak{n})$ and
$\kappa=S_{0}n_{i}E_{{\rm min},i}$ for $(q,n)=(1,2\mathfrak{n})$
lead apparently to the same $F_{\nu}$, while allowing the different interpretation
of the actual turbulent state of filaments and injection conditions.
At this juncture, here we demonstrate the specific calculation with such an invariant $\kappa$ 
and $(q,n)=(0,6)$ [$\leftrightarrow (1/3,7.2)$, $(1/2,8)$, etc.; 
the related coverage of $\lambda$ is verified below]. 
Indeed, the setting that tacitly allows for $n_{i}E_{{\rm min},i}\propto 1/S_{i}^{1-q}$ 
with $q<1$ is 
physically relevant in that the electron density is expected to be higher in the 
radially inner region of jet, where the electron 
energy to be higher, according to the X-ray observation suggesting that a large amount of 
energetic particles are confined in a narrower channel of the inner region  
\citep{2002ApJ...569...54K}. 

Figure \ref{fig:gamma-flux} shows, e.g. in knot A, the calculated $\gamma^{*}$ 
of electrons trapped and accelerated in the filaments of the various sizes ({\it top}) 
and the flux of synchrotron radiation from the electrons in some representative 
filaments ({\it bottom}). In the top panel, one can see that when the filament size 
$\lambda$ is larger than the critical size $\lambda_{\rm c}$, $\gamma^{*}$ is determined by the temporal limit due to 
the synchrotron cooling, while $\lambda<\lambda_{\rm c}$, $\gamma^{*}$ 
is determined by the spatial limit, as discussed in Sect. \ref{ssec:22}. 
The filaments 'a-d' 
labeled in the upper and lower panels correspond. At the largest filament 'a', 
the radio-optical flux is of the highest level owing to $B=B_{\rm m}$, while   
$\gamma^{*}$ is low. The cooling time $t_{\rm syn}\simeq 1.2\times 10^{8}$ s   
is much shorter than timescale of the escape from this filament, that is, of the order of $10^{10}$ s.
As for the thinner filament 'b', the flux decreases due to $B<B_{\rm m}$, and  
$\gamma^{*}$ increases. At the even thinner filament 'c' with $\lambda_{\rm c}=0.16$ pc, 
the flux drops further, whereas $\gamma^{*}$ and the corresponding synchrotron frequency reach the uppermost;   
$t_{\rm syn}\simeq 5.7\times 10^{8}\,{\rm s}$ is comparable to the escape time.
At 'd', where $\lambda$ and $B$ are still smaller, the escape from the filament dominates the synchrotron 
loss, and correspondingly, both $\gamma^{*}$ and the synchrotron frequency decrease
(thin dotted curve in the {\it bottom}). This figure 
also clarifies that synchrotron radiation from electrons in the thin filament of $\lambda\ll \lambda_{\rm c}$ 
contributes very little to the 'Total' spectrum, indicating that the coverage of $\lambda$ presumed 
above is reasonable. As a result, the radiation from the largest 'a' appears to form the break in the 
spectrum, while that from 'c', the cutoff of the spectrum. The synchrotron break and 
cutoff frequencies are defined by $\nu_{\rm b}=\nu_{0}(\gamma_{0}, B_{\rm m})$ and 
$\nu_{\rm c}=\nu(\gamma_{\rm c},B(\lambda=\lambda_{\rm c}))$, respectively, where $\gamma_{0}=\gamma^{*}(\lambda_{0})$ 
denoted as $\gamma_{\rm b}$, and $\gamma_{\rm c}=\gamma^{*}(\lambda=\lambda_{\rm c})$. 

Cooling after the acceleration, i.e., aging effect of electrons may be important for modeling a lower activity 
region similar to hot spots of Fanaroff-Riley class II (FR-II) radio galaxy lobes \citep[e.g., ][]{1997A&A...325...57M, 
2003MNRAS.345L..40B, 2005ApJ...628..104C}. The effect is, however, reasoned to be ignorable for the FR-I jet-knot environments,  
where the violent intermittent activities constantly take place so that a trail of working regions of the DSA 
successively captures electrons being radially confined by relatively ordered magnetic fields.  

\subsection{Inverse Comptonization}
\label{ssec:24}
For the major IC, we consider the {\it local} SSC. That is, the photon frequency boosted via the IC,    
$\nu_{\rm ic}=\gamma_{e}^{2}\nu$ being proportional to $\nu^{2}/B(\lambda_{i})$, is calculated for each 
filament; the corresponding photon fluxes from all filaments are 
summed up. Concerning $u_{\rm ph}$ as the target photon fields, 
$\langle u_{\rm ph}\rangle\lesssim \langle u_{\rm m}\rangle$ 
for the global mean over the concerned emission region is inferred from approximate equipartition relation 
inspected in an inner jet region closer to the nucleus \citep{2007A&A...471..453M} 
where the emission region be better matching 
electron acceleration region.  
In addition, in the situation in which filaments sparsely distribute (cf. Fig. \ref{fig:cutting-edge}),
is thought that the effect of radiative transfer 
is more pronounced in thinner filaments, causing larger deviation from the equipartition relation locally. 
Indeed, if all the low energy photons are ideally boosted up to higher energies by the IC,  
the evaluated flux of the high energy photons $(\nu F_{\nu})_{\rm ic}$ exceeds  
the observed level.   
In this aspect, we practically take account of the effect that prevents the ideal boosting of all 
photons.
First, the relation $u_{\rm ph}/u_{\rm m}\lesssim 1$ is expected to lower $(\nu F_{\nu})_{\rm ic}$ 
below the synchrotron $\nu F_{\nu}$-level. Second, for the thinner filaments, the larger 
emitting surface-to-volume ratio will get rise to more significant radiative cooling of 
filaments, namely, radiator effect. For one-dimensional approximation of a fragmented column, 
we employ the model that $u_{\rm ph}/u_{\rm m}$ decreases by the factor $\lambda/D$, 
instead of solving the details of the radiative transfer.
This would reflect the ratio of the scale $\lambda$ of filaments to
a typical mean free-path of synchrotron photons (close to the free-streaming limit, in cases).
Besides, the filament suffers synchrotron photons emanating from the other filaments.
And yet, even if energy loss of the photons during the inter-filament transport is little,
the influence of larger filaments on smaller ones is limited to the IC of the lower energy photons
(cf. Fig.~\ref{fig:gamma-flux} {\it bottom});
that of the smaller on the larger will be normally ignorable.
As is, the nonlocal SSC due to this effect is thought of as being less dominant,
compared with the local one.
Accordingly, combining 
the two, we have
$u_{\rm ph}/u_{\rm m}=a(\lambda/D)$ with $a\lesssim 1$, which is reflected in
$(\nu F_{\nu})_{\rm ic}/(\nu F_{\nu})_{\rm syn}$,
involving the implication of $a\sim f_{\rm cr}$ (cf. Table~1). 
It is noted that the model with $a<1$, substantiated below, ensures the maximum energy
analysis in which $t_{\rm acc}=t_{\rm syn}$ is solved for the electron DSA.
The Klein-Nishina effect is also included. 
Superposing the spectra that involve these effects, we obtain the boosted non-thermal spectrum 
as could be compared with the observed $\gamma$-ray fluxes. 

We shall mention, in addition to the synchrotron component, another potential candidate for target photons,
so-called external component: the contribution from the other parts of Cen A.
Considering the entire system, various regions are possible, such as the central engine, 
circumnuclear disk, thick dust lane, and so on. Of our particular interest is the region that may 
affect $\gamma$-ray hardening, and hence, we
focus on the nucleus with radiation contributions in the visible 
light and higher energy regions. The active nucleus was investigated by \citet{2000ApJ...528..276M} with 
WFPC2 and NICMOS observations of Hubble Space Telescope. They explain the optical and near-infrared emission by 
the combination of the non-thermal power-law extrapolated from X-rays reddened by $14~{\rm mag}$
and modified blackbody emission at 700 K probably caused by emission from hot dust within 2 pc of the nucleus. 
Alternatively, interferometric observations 
with MIDI revealed the existence of a well-resolved disk and an unresolved core in the innermost parsec of Cen A 
\citep{2007A&A...471..453M}. Using their results combined with the earlier data \citep{2000ApJ...528..276M}, they 
demonstrated that the core spectrum can be approximated by
the power-law synchrotron spectrum of $F_{\nu}\propto\nu^{-0.36}$ that is cut off exponentially above
the frequency of $\nu\sim 8\times 10^{13}~{\rm Hz}$, and the core is optically thick below $\sim 45~{\rm GHz}$.
Thermal component of geometrically thin disk whose diameter 
is about 0.6 pc containing dust heated up $\sim 240$ K also contributes between 20\%
(at the radio wavelength of $8~\mu{\rm m}$) and 40 \% (at $13~\mu{\rm m}$) to the nuclear flux. 
Even if the photons from the nuclear region could reach the knot A or B without significant energy loss, the 
Lorentz factor of the major electrons that would knock them up would be
of the order of $\gamma_{e}\sim 10^{4}-10^{5}$ (cf. Fig. 3 {\it top}).
Therefore, the frequency of boosted photons would be at most
$\nu_{\rm ic}=\gamma_{e}^{2}\nu_{\rm syn}\sim 10^{21}-10^{22}$ Hz, 
far below the H.E.S.S. band. The contribution of external components to the spectrum of the diffuse region 
will be discussed later.\\

\section{Results: Reproduction of the observed $\gamma$-ray fluxes}
\label{sec:3}
We apply the filament model discussed in Sect. \ref{ssec:21} to the knots A and B including 
many nested subknots and subsubknots, and determine the physical parameters required 
for the spectral evaluation. In addition to the key parameter $B_{\rm m}$,
the free dimensionless parameters of the model are $r$, $b$, $\beta$, $f_{\rm cr}$, and $a$, 
whose domains are $1<r\leq 4$, $b\ll 1$, $\beta\geq 2$, $f_{\rm cr}<1$, and 
$a\lesssim 1$, respectively. 
For the knots, it might seem difficult to fix the physical parameters because of lack 
of flux data in the infrared to ultraviolet band (Sect. \ref{sec:1}). However, we see that the values can be 
determined within a narrow allowable parameter-window, with reference to synchrotron flux 
data in the outer diffuse regions \citep{2006MNRAS.368L..15H} and $\gamma$-ray data 
\citep{2018A&A...619A..71H}.

In particular, the value of $B_{\rm m}$ is tightly restricted by the pivotal relation $\nu_{\rm ic,b}=
2.4\times 10^{23}(10\;{\rm mG}/B_{\rm m})(\nu_{\rm b}/10^{14}\;{\rm Hz})^{2}$, where $\nu_{\rm ic,b}$ 
is the frequency at which $(\nu F_{\nu})_{\rm ic}$-spectrum breaks such as to take the peak value.  
That is, we have $B_{\rm m}\sim10$ mG from combining the expected $\nu_{\rm ic,b}\gtrsim 10^{23}$ Hz 
and $\nu_{\rm b}\sim10^{14}$ Hz comparable to that observed in the outer diffuse regions. 
Regarding the shock strength, we suppose the situation in which the multiple shocks are not 
superposed, but spatially distributed, so that the compression ratio does not exceed 
the strong limit, as normal. 
Meanwhile, it is natural to consider that the shocks are weakened as they propagate outward; 
in the diffuse regions, radio-infrared data \citep{2006MNRAS.368L..15H} suggests 
$r\leq 3.3$. 
Accordingly, for the knots closer to the nucleus, the expected $r$-range reduces to $3.3\leq r\leq 4$. 
For given $D_{j}$, and the constraints on $B_{\rm m}$ and $r$, we can estimate the value 
of $b$ from 
\begin{equation}
b\simeq 0.055 \frac{r}{r-1}f(r,\theta)\left(\frac{\nu_{\rm b}}{10^{14}\;{\rm Hz}}\right)^{\frac{2}{3}}
\left(\frac{D}{100\;{\rm pc}}\right)^{\frac{2}{3}}\left(\frac{B_{\rm m}}{10\;{\rm mG}}\right),
\label{eqn:b}
\end{equation}
where $f(r,\theta)$ is the factor $[1+\cdots]$ given in Eq. (\ref{eqn:eta}).
Provided the $\nu_{\rm b}\sim 10^{14}$ Hz and the plausible range of $f_{\rm cr}\gtrsim 10^{-1}$, 
therefore, we have $b\sim 10^{-1}$, i.e., virtually the maximum. 
As concerns $\beta$, it is unlikely that the effective value largely deviates from 2 
(cf. Sect. \ref{ssec:21}). 
And, its combination with the $r$-value determines X-ray spectral index. 
Following this, specific shape of the $\gamma$-ray spectrum is determined, whereupon the 
flux level is adjusted by the coefficient $a$ of the cooling factor. 
Assuming that absorption of the $\gamma$-rays by diffuse extragalactic 
background light is negligible \citep{Aharonian04}, the theoretical spectrum is constructed 
as better fit to the fluxes measured in multi-band over radio-to-VHE $\gamma$-rays. 

\begin{table}\label{tab:knot_prop}
  \caption{Physical properties of the major knots and diffuse region.}{%
  \begin{tabular}{lccccccc} 
      \hline
      Name\tablefootmark{a} & $D_{j}$\tablefootmark{b} (pc) & $B_{\rm m}$ (G) & $r$ & $b$ & $\beta$ & $f_{\rm cr}$ & $a$ \\
      \hline
      knot A & 147.9 & 0.01 & 3.5 & 0.1 & 3 & 0.5 & 0.4\\
      knot B & 173.4 & 0.01 & 3.4 & 0.2 & 3 & 0.5 & 0.4\\
      Diffuse  & 714 & 0.01 & 3.3 & 0.3 & $-$ & 0.1 & 0.4\\
      \hline
    \end{tabular}}
\tablefoot{
\tablefoottext{a}{Name of the X-ray emitting regions that contain smaller X-ray (sub-/subsub-)knots. 
	The diffuse region corresponds to the rectangle region 'Inner' defined in 
	\citet{2006MNRAS.368L..15H}.}
\tablefoottext{b}{The jet width at each region (for the knots, radio-20 cm: 
        BFS83).}
}
\end{table}

\begin{figure}[t]
\begin{center}
\includegraphics[trim = 0 0 0 0, width=8cm,clip]{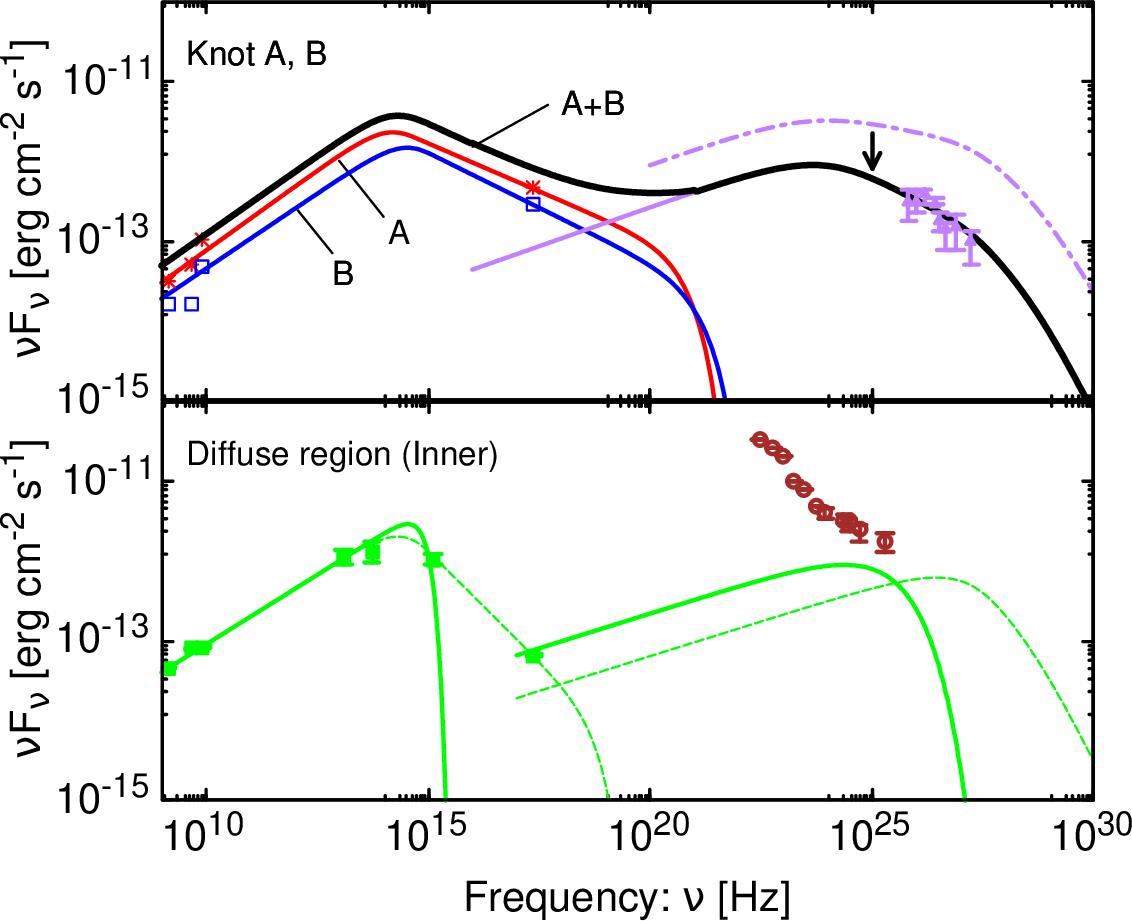} 
\end{center}
\caption{The $\nu F_{\nu}$-spectra of the knots ({\it top}) and diffuse region ({\it bottom}) 
    in the range of radio to $\gamma$-rays, for the parameter values given in Table 1.  
    For the knots, the observed radio fluxes are cited from BFS83, and the  X-ray 
    fluxes are obtained by adding up the fluxes of subknots \citep{2002ApJ...569...54K}. 
    The calculated spectra and observed fluxes of knot A/B are denoted by red/blue curves and marks 
    (asterisks/open-squares), respectively. The 
    purple solid and dot-dashed curves show the calculated spectra of the IC component 
    (knot A+B) with and without radiative cooling effect of filaments, respectively. The 
    overall spectrum of synchrotron+IC is shown as the black solid curve. The calculated spectra 
    of the diffuse region are denoted by green curves, and the observed fluxes from that region (green crosses) 
    are cited from \citet{2006MNRAS.368L..15H}. The cases for which the observed 
    X-ray flux is assumed to originate from synchrotron radiation and IC(+radiative cooling) are respectively shown 
    by the thin dashed and bold curves. The observed $\gamma$-ray fluxes of the H.E.S.S. ({\it top}: purple triangles) and 
    Fermi-LAT ({\it bottom}: brown open-circles) are cited from \citet{2018A&A...619A..71H}.}
\label{fig:spect_cenaabf} 
\end{figure}

The parameter values we have used here are summarized in Table 1,  
along with the observed $D_{j}$ of the major knots and the diffuse region. 
For the knots A and B, the synchrotron spectra 
are shown in Fig. \ref{fig:spect_cenaabf} ({\it top}) with red and blue curves, respectively; the former 
corresponds to 'Total' in Fig. \ref{fig:gamma-flux} ({\it bottom}).  
For the knot A (B), the X-ray $F_{\nu}$-spectral index is estimated 1.2
(1.3), and $\gamma_{e}$ of the electrons making the major contribution to the
spectrum at, e.g., $1.4~{\rm keV}$ is $1.3\,(1.1)\times 10^{7}$,
generated in the size of $\lambda=3.5\,(5.8)~{\rm pc}$ where $B=0.48\,(0.67)~{\rm mG}$. 
For convenience, 
the resulting values of $\gamma_{\rm b}$, $\gamma_{\rm c}$, $\nu_{\rm b}$, and $\nu_{\rm c}$  
are listed in Table 2 
(for knot A, compare Fig. \ref{fig:gamma-flux}). 
The $\gamma_{\rm c}\sim 10^{9}$ provides the 
theoretical upper limit of the electron Lorentz factor, as amenable to   
the argument in \citet{2020Natur.582..356H}. 
We find that $\nu_{\rm c}$ reaches MeV $\gamma$-ray band and the spectra extended above $\nu_{\rm b}$ 
in the bright knots make dominant contribution to establishing the 
observed $\gamma$-ray hardening. 
In the {\it top} panel, the theoretical $(\nu F_{\nu})_{\rm ic}$-spectra
including the Klein-Nishina (KN) effect and the KN plus radiator effects are
denoted by dot-dashed and solid curves, respectively.  
The spectral depression by the radiative cooling is indicated by the arrow, and 
found to be noticeable in the VHE range and above.
The overall multiwavelength spectrum of the combined synchrotron and IC components is
also shown by the thick black solid. 

\begin{table}\label{tab:knot_freq}
  \caption{Break and cutoff frequencies and the corresponding Lorentz factor of electron.}{%
  \begin{tabular}{lcccc} 
      \hline
      Name & $\gamma_{\rm b}$ & $\gamma_{\rm c}$ & $\nu_{\rm b}$ (Hz) & $\nu_{\rm c}$ (Hz) \\
      \hline
      knot A & $6.3\times 10^{4}$ & $3.0\times 10^{9}$ 
      & $1.7\times 10^{14}$ & $8.2\times 10^{20}$ \\
      knot B & $9.7\times 10^{4}$ & $4.0\times 10^{9}$ 
      & $3.9\times 10^{14}$ & $1.6\times 10^{21}$ \\
      Inner  & $-$ & $1.4\times 10^{5}$ & $-$ & $8.5\times 10^{14}$\\
      \hline
    \end{tabular}}
\end{table}

Some lower luminosity X-ray features \citep[e.g., FX1, $\cdots$, GX1, $\cdots$: ][]{2019ApJ...871..248S} that have 
recently been resolved in the outer diffuse regions \citep['Inner' and 'Middle': ][]{2006MNRAS.368L..15H}  
appear to be similar to the well-known X-ray subknots \citep[AX1, $\cdots$, BX1, $\cdots$: ][]{2002ApJ...569...54K}  
in the knots (A and B: BFS83). This potentially allows us to apply the current method 
to the diffuse regions in a common way, except for differences in the physical parameters. 
In Fig. \ref{fig:spect_cenaabf} ({\it bottom}), for the parameters of the diffuse region given in Table 1, 
with $\beta=6$, we show the spectrum of synchrotron radiation from the region  
'Inner' and its upscattered spectrum via the IC, by the green dashed curves. We see that interpreting the 
observed X-ray flux as the synchrotron origin fails in reproducing the $\gamma$-ray data, as far as the 
SSC is presumed. It does not mean, however, that the filament model itself is ruled out there. 
Coupling with the external Comptonization of dust emissions (from the nuclear disk or dust lane as 
mentioned above), one might account for the observed $\gamma$-ray spectrum up to the VHE-range. However, here we 
consider an alternative  
possibility: the case in which filamentation has already ceased and coalescence of filaments has proceeded, 
to leave a single large filament, along the discussion in Sect. \ref{ssec:21}. This 
just corresponds to the single-zone modeling; for the same 
parameter set (but, now free from $\beta$), we show the spectrum 
in the same panel ({\it bottom}) by the thick green solid. For this case, the values of $\gamma_{\rm c}(=\gamma_{\rm b})$ 
and $\nu_{\rm c}(=\nu_{\rm b})$ are listed in Table 2. 
Although the solo-contribution from the 'Inner' part of the diffuse region is not quite enough to account for 
the observed Fermi fluxes, it is anticipated that the observed ones be explained by adding the contributions 
from the knots and/or the other diffuse regions. 

\section{Concluding remarks}
\label{sec:4}
In conclusion, we found that major origin of the spectral hardening of the VHE 
$\gamma$-ray fluxes in Cen A could be ascribed to the inhomogeneous SSC in the bright X-ray knots 
distributed in the jet; synchrotron loss of the DSA is markedly suppressed for the electrons trapped in the weak 
magnetic field of thin filaments. In particular, the model reflects 
the common self-similar structure and turbulent evolution of the internal jet, and as is, provides a promising 
alternative to the external Compton \citep{2019ApJ...878..139T} and shear acceleration 
\citep{2017ApJ...842...39L} scenarios. 
In our concept, once the filament parameters, which are somewhat constrained 
by independent observations, 
are fixed, the key factors characterizing the spectral shape (e.g., the break and cutoff frequencies, the 
power-law exponent, etc.) are derived. Unlike conventional single-zone SSC models, it is no 
longer necessary to explicitly specify the maximum and minimum values of the electron Lorentz factor or the 
2nd exponent of the broken power-law spectrum. 

This paper does not deal with the steep rise in 
$\nu F_{\nu}$-flux observed by the TANAMI VLBI \citep[also, from the archival data:][]
{2010ApJ...719.1433A,2000ApJ...528..276M}, which is deemed to be the thermal 
and/or non-thermal component 
ranging from radio to infrared wavelengths. The narrow high energy peak between the 
X- and soft $\gamma$-rays observed by SUZAKU \citep{2007ApJ...665..209M} and SWIFT-BAT 
\citep{2008ApJ...673...96A}, respectively (not shown in the figure), which is thought of 
as comprising the 
upscattered thermal photons \citep[or the synchrotron photons;][]{2008A&A...478..111L}, 
is also excluded from the scope of this study. 

As for the spectral variability of Cen A, stable nature of the X-ray spectrum was first revealed by 
Rossi X-ray Timing Explorer (RXTE) observations \citep{2011ApJ...733...23R}. On the other hand, 
\citet{2016PASJ...68S..25T} reported a soft-lag in X-ray band on the timescale of days, owing to detailed 
analysis of the light curves measured by the GSC aboard the Monitor of 
All-sky X-ray Image (MAXI). Although the mechanism that has generated the soft-lag is still 
unknown, the results are negative for a single-zone jet, incorporated in SSC model. 
According to our model, the soft-lag can be explained by the stronger turbulent state in an innermost 
region (compared to compact blazars), which reduces the $\beta$-value to $\sim 2$, 
or even less \citep{2008ApJ...675L..61H}.
Our SSC model can also respond to variability of $\gamma$-rays likewise that of synchrotron 
X-rays; that is, the $\gamma$-ray variability timescale can be estimated as 
$\tau\sim t_{\rm acc}=t_{\rm syn}\propto\lambda^{-(3\beta-7)/8}$, with $\lambda$ being the solution of 
$\nu_{\gamma}=\gamma_{e}^{2}\nu(\gamma_{e},B)$ for given $\gamma$-ray energy 
$h\nu_{\gamma}=E$. Note that for $\beta>7/3$ hard-lag appears. At $E=20$ GeV (100 TeV) the Cherenkov Telescope Array 
\citep[CTA; ][]{2020NatAs...4..124B} covers, we read $\tau\sim 4.2$ yr (6.0 yr) for, 
e.g., the knot A parameters,  
along with the hard-lag appearance above GeV. We hope that the filament model, and its 
applicability to the other jets, be verified by high spatiotemporal resolution, 
multiwavelength observations including the CTA in future. \\

\bibliographystyle{aa}
\bibliography{main}

@ARTICLE{2018A&A...619A..71H,
       author = {{Abdalla}, H. and {Abramowski}, A. and {Aharonian}, F. and {Ait Benkhali}, F. and {Ang{\"u}ner}, E.~O. and {Arakawa}, M. and {Armand}, C. and {Arrieta}, M. and {Backes}, M. and {Balzer}, A. and {Barnard}, M. and {Becherini}, Y. and {Becker Tjus}, J. and {Berge}, D. and {Bernhard}, S. and {Bernl{\"o}hr}, K. and {Blackwell}, R. and {B{\"o}ttcher}, M. and {Boisson}, C. and {Bolmont}, J. and {Bonnefoy}, S. and {Bordas}, P. and {Bregeon}, J. and {Brun}, F. and {Brun}, P. and {Bryan}, M. and {B{\"u}chele}, M. and {Bulik}, T. and {Capasso}, M. and {Caroff}, S. and {Carosi}, A. and {Casanova}, S. and {Cerruti}, M. and {Chakraborty}, N. and {Chaves}, R.~C.~G. and {Chen}, A. and {Chevalier}, J. and {Colafrancesco}, S. and {Condon}, B. and {Conrad}, J. and {Davids}, I.~D. and {Decock}, J. and {Deil}, C. and {Devin}, J. and {deWilt}, P. and {Dirson}, L. and {Djannati-Ata{\"\i}}, A. and {Donath}, A. and {Drury}, L.~O. 'C. and {Dyks}, J. and {Edwards}, T. and {Egberts}, K. and {Emery}, G. and {Ernenwein}, J. -P. and {Eschbach}, S. and {Farnier}, C. and {Fegan}, S. and {Fernandes}, M.~V. and {Fiasson}, A. and {Fontaine}, G. and {Funk}, S. and {F{\"u}{\ss}ling}, M. and {Gabici}, S. and {Gallant}, Y.~A. and {Garrigoux}, T. and {Gat{\'e}}, F. and {Giavitto}, G. and {Glawion}, D. and {Glicenstein}, J.~F. and {Gottschall}, D. and {Grondin}, M. -H. and {Hahn}, J. and {Haupt}, M. and {Hawkes}, J. and {Heinzelmann}, G. and {Henri}, G. and {Hermann}, G. and {Hinton}, J.~A. and {Hofmann}, W. and {Hoischen}, C. and {Holch}, T.~L. and {Holler}, M. and {Horns}, D. and {Ivascenko}, A. and {Iwasaki}, H. and {Jacholkowska}, A. and {Jamrozy}, M. and {Jankowsky}, D. and {Jankowsky}, F. and {Jingo}, M. and {Jouvin}, L. and {Jung-Richardt}, I. and {Kastendieck}, M.~A. and {Katarzy{\'n}ski}, K. and {Katsuragawa}, M. and {Katz}, U. and {Kerszberg}, D. and {Khangulyan}, D. and {Kh{\'e}lifi}, B. and {King}, J. and {Klepser}, S. and {Klochkov}, D. and {Klu{\'z}niak}, W. and {Komin}, Nu. and {Kosack}, K. and {Krakau}, S. and {Kraus}, M. and {Kr{\"u}ger}, P.~P. and {Laffon}, H. and {Lamanna}, G. and {Lau}, J. and {Lefaucheur}, J. and {Lemi{\`e}re}, A. and {Lemoine-Goumard}, M. and {Lenain}, J. -P. and {Leser}, E. and {Lohse}, T. and {Lorentz}, M. and {Liu}, R. and {L{\'o}pez-Coto}, R. and {Lypova}, I. and {Malyshev}, D. and {Marandon}, V. and {Marcowith}, A. and {Mariaud}, C. and {Marx}, R. and {Maurin}, G. and {Maxted}, N. and {Mayer}, M. and {Meintjes}, P.~J. and {Meyer}, M. and {Mitchell}, A.~M.~W. and {Moderski}, R. and {Mohamed}, M. and {Mohrmann}, L. and {Mor{\r{a}}}, K. and {Moulin}, E. and {Murach}, T. and {Nakashima}, S. and {de Naurois}, M. and {Ndiyavala}, H. and {Niederwanger}, F. and {Niemiec}, J. and {Oakes}, L. and {O'Brien}, P. and {Odaka}, H. and {Ohm}, S. and {Ostrowski}, M. and {Oya}, I. and {Padovani}, M. and {Panter}, M. and {Parsons}, R.~D. and {Pekeur}, N.~W. and {Pelletier}, G. and {Perennes}, C. and {Petrucci}, P. -O. and {Peyaud}, B. and {Piel}, Q. and {Pita}, S. and {Poireau}, V. and {Prokhorov}, D.~A. and {Prokoph}, H. and {P{\"u}hlhofer}, G. and {Punch}, M. and {Quirrenbach}, A. and {Raab}, S. and {Rauth}, R. and {Reimer}, A. and {Reimer}, O. and {Renaud}, M. and {de los Reyes}, R. and {Rieger}, F. and {Rinchiuso}, L. and {Romoli}, C. and {Rowell}, G. and {Rudak}, B. and {Rulten}, C.~B. and {Sahakian}, V. and {Saito}, S. and {Sanchez}, D.~A. and {Santangelo}, A. and {Sasaki}, M. and {Schlickeiser}, R. and {Sch{\"u}ssler}, F. and {Schulz}, A. and {Schwanke}, U. and {Schwemmer}, S. and {Seglar-Arroyo}, M. and {Seyffert}, A.~S. and {Shafi}, N. and {Shilon}, I. and {Shiningayamwe}, K. and {Simoni}, R. and {Sol}, H. and {Spanier}, F. and {Spir-Jacob}, M. and {Stawarz}, {\L}. and {Steenkamp}, R. and {Stegmann}, C. and {Steppa}, C. and {Sushch}, I. and {Takahashi}, T. and {Tavernet}, J. -P. and {Tavernier}, T. and {Taylor}, A.~M. and {Terrier}, R. and {Tibaldo}, L. and {Tiziani}, D. and {Tluczykont}, M. and {Trichard}, C. and {Tsirou}, M. and {Tsuji}, N. and {Tuffs}, R. and {Uchiyama}, Y. and {van der Walt}, D.~J. and {van Eldik}, C. and {van Rensburg}, C. and {van Soelen}, B. and {Vasileiadis}, G. and {Veh}, J. and {Venter}, C. and {Viana}, A. and {Vincent}, P. and {Vink}, J. and {Voisin}, F. and {V{\"o}lk}, H.~J. and {Vuillaume}, T. and {Wadiasingh}, Z. and {Wagner}, S.~J. and {Wagner}, P. and {Wagner}, R.~M. and {White}, R. and {Wierzcholska}, A. and {Willmann}, P. and {W{\"o}rnlein}, A. and {Wouters}, D. and {Yang}, R. and {Zaborov}, D. and {Zacharias}, M. and {Zanin}, R. and {Zdziarski}, A.~A. and {Zech}, A. and {Zefi}, F. and {Ziegler}, A. and {Zorn}, J. and {{\.Z}ywucka}, N. and {Fermi-LAT Collaboration} and {Magill}, J.~D. and {Buson}, S. and {Cheung}, C.~C. and {Perkins}, J.~S. and {Tanaka}, Y.},
        title = "{The {\ensuremath{\gamma}}-ray spectrum of the core of Centaurus A as observed with H.E.S.S. and Fermi-LAT}",
      journal = {\aap},
         year = 2018,
        month = nov,
       volume = {619},
          eid = {A71},
        pages = {A71},
          doi = {10.1051/0004-6361/201832640},
archivePrefix = {arXiv},
       eprint = {1807.07375},
 primaryClass = {astro-ph.HE},
       adsurl = {https://ui.adsabs.harvard.edu/abs/2018A&A...619A..71H}
}

@ARTICLE{2020Natur.582..356H,
       author = {{Abdalla}, H. and {Adam}, R. and {Aharonian}, F. and {Ait Benkhali}, F. and {Ang{\"u}ner}, E.~O. and {Arakawa}, M. and {Arcaro}, C. and {Armand}, C. and {Ashkar}, H. and {Backes}, M. and {Barbosa Martins}, V. and {Barnard}, M. and {Becherini}, Y. and {Berge}, D. and {Bernl{\"o}hr}, K. and {Blackwell}, R. and {B{\"o}ttcher}, M. and {Boisson}, C. and {Bolmont}, J. and {Bonnefoy}, S. and {Bregeon}, J. and {Breuhaus}, M. and {Brun}, F. and {Brun}, P. and {Bryan}, M. and {B{\"u}chele}, M. and {Bulik}, T. and {Bylund}, T. and {Capasso}, M. and {Caroff}, S. and {Carosi}, A. and {Casanova}, S. and {Cerruti}, M. and {Chand}, T. and {Chandra}, S. and {Chen}, A. and {Colafrancesco}, S. and {Cury{\l}o}, M. and {Davids}, I.~D. and {Deil}, C. and {Devin}, J. and {deWilt}, P. and {Dirson}, L. and {Djannati-Ata{\"\i}}, A. and {Dmytriiev}, A. and {Donath}, A. and {Doroshenko}, V. and {Drury}, L. O'C. and {Dyks}, J. and {Egberts}, K. and {Emery}, G. and {Ernenwein}, J. -P. and {Eschbach}, S. and {Feijen}, K. and {Fegan}, S. and {Fiasson}, A. and {Fontaine}, G. and {Funk}, S. and {F{\"u}{\ss}ling}, M. and {Gabici}, S. and {Gallant}, Y.~A. and {Gat{\'e}}, F. and {Giavitto}, G. and {Glawion}, D. and {Glicenstein}, J.~F. and {Gottschall}, D. and {Grondin}, M. -H. and {Hahn}, J. and {Haupt}, M. and {Heinzelmann}, G. and {Henri}, G. and {Hermann}, G. and {Hinton}, J.~A. and {Hofmann}, W. and {Hoischen}, C. and {Holch}, T.~L. and {Holler}, M. and {Horns}, D. and {Huber}, D. and {Iwasaki}, H. and {Jamrozy}, M. and {Jankowsky}, D. and {Jankowsky}, F. and {Jardin-Blicq}, A. and {Jung-Richardt}, I. and {Kastendieck}, M.~A. and {Katarzy{\'n}ski}, K. and {Katsuragawa}, M. and {Katz}, U. and {Khangulyan}, D. and {Kh{\'e}lifi}, B. and {King}, J. and {Klepser}, S. and {Klu{\'z}niak}, W. and {Komin}, N. and {Kosack}, K. and {Kostunin}, D. and {Kraus}, M. and {Lamanna}, G. and {Lau}, J. and {Lemi{\`e}re}, A. and {Lemoine-Goumard}, M. and {Lenain}, J. -P. and {Leser}, E. and {Levy}, C. and {Lohse}, T. and {Lypova}, I. and {Mackey}, J. and {Majumdar}, J. and {Malyshev}, D. and {Marandon}, V. and {Marcowith}, A. and {Mares}, A. and {Mariaud}, C. and {Mart{\'\i}-Devesa}, G. and {Marx}, R. and {Maurin}, G. and {Meintjes}, P.~J. and {Mitchell}, A.~M.~W. and {Moderski}, R. and {Mohamed}, M. and {Mohrmann}, L. and {Moore}, C. and {Moulin}, E. and {Muller}, J. and {Murach}, T. and {Nakashima}, S. and {de Naurois}, M. and {Ndiyavala}, H. and {Niederwanger}, F. and {Niemiec}, J. and {Oakes}, L. and {O'Brien}, P. and {Odaka}, H. and {Ohm}, S. and {de Ona Wilhelmi}, E. and {Ostrowski}, M. and {Oya}, I. and {Panter}, M. and {Parsons}, R.~D. and {Perennes}, C. and {Petrucci}, P. -O. and {Peyaud}, B. and {Piel}, Q. and {Pita}, S. and {Poireau}, V. and {Priyana Noel}, A. and {Prokhorov}, D.~A. and {Prokoph}, H. and {P{\"u}hlhofer}, G. and {Punch}, M. and {Quirrenbach}, A. and {Raab}, S. and {Rauth}, R. and {Reimer}, A. and {Reimer}, O. and {Remy}, Q. and {Renaud}, M. and {Rieger}, F. and {Rinchiuso}, L. and {Romoli}, C. and {Rowell}, G. and {Rudak}, B. and {Ruiz-Velasco}, E. and {Sahakian}, V. and {Saito}, S. and {Sanchez}, D.~A. and {Santangelo}, A. and {Sasaki}, M. and {Schlickeiser}, R. and {Sch{\"u}ssler}, F. and {Schulz}, A. and {Schutte}, H.~M. and {Schwanke}, U. and {Schwemmer}, S. and {Seglar-Arroyo}, M. and {Senniappan}, M. and {Seyffert}, A.~S. and {Shafi}, N. and {Shiningayamwe}, K. and {Simoni}, R. and {Sinha}, A. and {Sol}, H. and {Specovius}, A. and {Spir-Jacob}, M. and {Stawarz}, {\L}. and {Steenkamp}, R. and {Stegmann}, C. and {Steppa}, C. and {Takahashi}, T. and {Tavernier}, T. and {Taylor}, A.~M. and {Terrier}, R. and {Tiziani}, D. and {Tluczykont}, M. and {Trichard}, C. and {Tsirou}, M. and {Tsuji}, N. and {Tuffs}, R. and {Uchiyama}, Y. and {van der Walt}, D.~J. and {van Eldik}, C. and {van Rensburg}, C. and {van Soelen}, B. and {Vasileiadis}, G. and {Veh}, J. and {Venter}, C. and {Vincent}, P. and {Vink}, J. and {Voisin}, F. and {V{\"o}lk}, H.~J. and {Vuillaume}, T. and {Wadiasingh}, Z. and {Wagner}, S.~J. and {White}, R. and {Wierzcholska}, A. and {Yang}, R. and {Yoneda}, H. and {Zacharias}, M. and {Zanin}, R. and {Zdziarski}, A.~A. and {Zech}, A. and {Ziegler}, A. and {Zorn}, J. and {{\.Z}ywucka}, N.},
        title = "{Resolving acceleration to very high energies along the jet of Centaurus A}",
      journal = {\nat},
         year = 2020,
        month = jun,
       volume = {582},
       number = {7812},
        pages = {356-359},
          doi = {10.1038/s41586-020-2354-1},
archivePrefix = {arXiv},
       eprint = {2007.04823},
 primaryClass = {astro-ph.HE},
       adsurl = {https://ui.adsabs.harvard.edu/abs/2020Natur.582..356H}
}

@ARTICLE{2010ApJ...719.1433A,
       author = {{Abdo}, A.~A. and {Ackermann}, M. and {Ajello}, M. and {Atwood}, W.~B. and {Baldini}, L. and {Ballet}, J. and {Barbiellini}, G. and {Bastieri}, D. and {Baughman}, B.~M. and {Bechtol}, K. and {Bellazzini}, R. and {Berenji}, B. and {Blandford}, R.~D. and {Bloom}, E.~D. and {Bonamente}, E. and {Borgland}, A.~W. and {Bouvier}, A. and {Brandt}, T.~J. and {Bregeon}, J. and {Brez}, A. and {Brigida}, M. and {Bruel}, P. and {Buehler}, R. and {Buson}, S. and {Caliandro}, G.~A. and {Cameron}, R.~A. and {Cannon}, A. and {Caraveo}, P.~A. and {Carrigan}, S. and {Casandjian}, J.~M. and {Cavazzuti}, E. and {Cecchi}, C. and {{\c{C}}elik}, {\"O}. and {Charles}, E. and {Chekhtman}, A. and {Cheung}, C.~C. and {Chiang}, J. and {Ciprini}, S. and {Claus}, R. and {Cohen-Tanugi}, J. and {Colafrancesco}, S. and {Cominsky}, L.~R. and {Conrad}, J. and {Costamante}, L. and {Davis}, D.~S. and {Dermer}, C.~D. and {de Angelis}, A. and {de Palma}, F. and {Silva}, E. do Couto e. and {Drell}, P.~S. and {Dubois}, R. and {Dumora}, D. and {Falcone}, A. and {Farnier}, C. and {Favuzzi}, C. and {Fegan}, S.~J. and {Finke}, J. and {Focke}, W.~B. and {Fortin}, P. and {Frailis}, M. and {Fukazawa}, Y. and {Funk}, S. and {Fusco}, P. and {Gargano}, F. and {Gasparrini}, D. and {Gehrels}, N. and {Georganopoulos}, M. and {Germani}, S. and {Giebels}, B. and {Giglietto}, N. and {Giommi}, P. and {Giordano}, F. and {Giroletti}, M. and {Glanzman}, T. and {Godfrey}, G. and {Grandi}, P. and {Grenier}, I.~A. and {Grondin}, M. -H. and {Grove}, J.~E. and {Guillemot}, L. and {Guiriec}, S. and {Hadasch}, D. and {Harding}, A.~K. and {Hase}, Hayo and {Hayashida}, M. and {Hays}, E. and {Horan}, D. and {Hughes}, R.~E. and {Itoh}, R. and {Jackson}, M.~S. and {J{\'o}hannesson}, G. and {Johnson}, A.~S. and {Johnson}, T.~J. and {Johnson}, W.~N. and {Kadler}, M. and {Kamae}, T. and {Katagiri}, H. and {Kataoka}, J. and {Kawai}, N. and {Kishishita}, T. and {Kn{\"o}dlseder}, J. and {Kuss}, M. and {Lande}, J. and {Latronico}, L. and {Lee}, S. -H. and {Lemoine-Goumard}, M. and {Llena Garde}, M. and {Longo}, F. and {Loparco}, F. and {Lott}, B. and {Lovellette}, M.~N. and {Lubrano}, P. and {Makeev}, A. and {Mazziotta}, M.~N. and {McConville}, W. and {McEnery}, J.~E. and {Michelson}, P.~F. and {Mitthumsiri}, W. and {Mizuno}, T. and {Moiseev}, A.~A. and {Monte}, C. and {Monzani}, M.~E. and {Morselli}, A. and {Moskalenko}, I.~V. and {Murgia}, S. and {M{\"u}ller}, C. and {Nakamori}, T. and {Naumann-Godo}, M. and {Nolan}, P.~L. and {Norris}, J.~P. and {Nuss}, E. and {Ohno}, M. and {Ohsugi}, T. and {Ojha}, R. and {Okumura}, A. and {Omodei}, N. and {Orlando}, E. and {Ormes}, J.~F. and {Ozaki}, M. and {Pagani}, C. and {Paneque}, D. and {Panetta}, J.~H. and {Parent}, D. and {Pelassa}, V. and {Pepe}, M. and {Pesce-Rollins}, M. and {Piron}, F. and {Pl{\"o}tz}, C. and {Porter}, T.~A. and {Rain{\`o}}, S. and {Rando}, R. and {Razzano}, M. and {Razzaque}, S. and {Reimer}, A. and {Reimer}, O. and {Reposeur}, T. and {Ripken}, J. and {Ritz}, S. and {Rodriguez}, A.~Y. and {Roth}, M. and {Ryde}, F. and {Sadrozinski}, H.~F. -W. and {Sanchez}, D. and {Sander}, A. and {Scargle}, J.~D. and {Sgr{\`o}}, C. and {Siskind}, E.~J. and {Smith}, P.~D. and {Spandre}, G. and {Spinelli}, P. and {Starck}, J. -L. and {Stawarz}, L. and {Strickman}, M.~S. and {Suson}, D.~J. and {Tajima}, H. and {Takahashi}, H. and {Takahashi}, T. and {Tanaka}, T. and {Thayer}, J.~B. and {Thayer}, J.~G. and {Thompson}, D.~J. and {Tibaldo}, L. and {Torres}, D.~F. and {Tosti}, G. and {Tramacere}, A. and {Uchiyama}, Y. and {Usher}, T.~L. and {Vandenbroucke}, J. and {Vasileiou}, V. and {Vilchez}, N. and {Vitale}, V. and {Waite}, A.~P. and {Wang}, P. and {Winer}, B.~L. and {Wood}, K.~S. and {Yang}, Z. and {Ylinen}, T. and {Ziegler}, M.},
        title = "{Fermi Large Area Telescope View of the Core of the Radio Galaxy Centaurus A}",
      journal = {\apj},
         year = 2010,
        month = aug,
       volume = {719},
       number = {2},
        pages = {1433-1444},
          doi = {10.1088/0004-637X/719/2/1433},
archivePrefix = {arXiv},
       eprint = {1006.5463},
 primaryClass = {astro-ph.HE},
       adsurl = {https://ui.adsabs.harvard.edu/abs/2010ApJ...719.1433A}
}

@BOOK{aharonian04,
       author = {{Aharonian}, F.~A.},
        title = "{Very High Energy Cosmic Gamma Radiation: A Crucial Window on the Extreme Universe}",
    publisher = {World Scientific Publishing},
	 year = "2004",
      edition = {1},
	 isbn = {978-981-02-4573-3}
}

@ARTICLE{2009ApJ...695L..40A,
       author = {{Aharonian}, F. and {Akhperjanian}, A.~G. and {Anton}, G. and {de Almeida}, U. Barres and {Bazer-Bachi}, A.~R. and {Becherini}, Y. and {Behera}, B. and {Benbow}, W. and {Bernl{\"o}hr}, K. and {Boisson}, C. and {Bochow}, A. and {Borrel}, V. and {Brion}, E. and {Brucker}, J. and {Brun}, P. and {B{\"u}hler}, R. and {Bulik}, T. and {B{\"u}sching}, I. and {Boutelier}, T. and {Chadwick}, P.~M. and {Charbonnier}, A. and {Chaves}, R.~C.~G. and {Cheesebrough}, A. and {Chounet}, L. -M. and {Clapson}, A.~C. and {Coignet}, G. and {Dalton}, M. and {Daniel}, M.~K. and {Davids}, I.~D. and {Degrange}, B. and {Deil}, C. and {Dickinson}, H.~J. and {Djannati-Ata{\"\i}}, A. and {Domainko}, W. and {Drury}, L. O'C. and {Dubois}, F. and {Dubus}, G. and {Dyks}, J. and {Dyrda}, M. and {Egberts}, K. and {Emmanoulopoulos}, D. and {Espigat}, P. and {Farnier}, C. and {Feinstein}, F. and {Fiasson}, A. and {F{\"o}rster}, A. and {Fontaine}, G. and {F{\"u}{\ss}ling}, M. and {Gabici}, S. and {Gallant}, Y.~A. and {G{\'e}rard}, L. and {Giebels}, B. and {Glicenstein}, J. -F. and {Gl{\"u}ck}, B. and {Goret}, P. and {G{\"o}hring}, D. and {Hauser}, D. and {Hauser}, M. and {Heinz}, S. and {Heinzelmann}, G. and {Henri}, G. and {Hermann}, G. and {Hinton}, J.~A. and {Hoffmann}, A. and {Hofmann}, W. and {Holleran}, M. and {Hoppe}, S. and {Horns}, D. and {Jacholkowska}, A. and {de Jager}, O.~C. and {Jahn}, C. and {Jung}, I. and {Katarzy{\'n}ski}, K. and {Katz}, U. and {Kaufmann}, S. and {Kendziorra}, E. and {Kerschhaggl}, M. and {Khangulyan}, D. and {Kh{\'e}lifi}, B. and {Keogh}, D. and {Klu{\'z}niak}, W. and {Kneiske}, T. and {Komin}, Nu. and {Kosack}, K. and {Lamanna}, G. and {Latham}, I.~J. and {Lenain}, J. -P. and {Lohse}, T. and {Marandon}, V. and {Martin}, J.~M. and {Martineau-Huynh}, O. and {Marcowith}, A. and {Maurin}, D. and {McComb}, T.~J.~L. and {Medina}, M.~C. and {Moderski}, R. and {Moulin}, E. and {Naumann-Godo}, M. and {de Naurois}, M. and {Nedbal}, D. and {Nekrassov}, D. and {Niemiec}, J. and {Nolan}, S.~J. and {Ohm}, S. and {Olive}, J. -F. and {de O{\~n}a Wilhelmi}, E. and {Orford}, K.~J. and {Ostrowski}, M. and {Panter}, M. and {Arribas}, M. Paz and {Pedaletti}, G. and {Pelletier}, G. and {Petrucci}, P. -O. and {Pita}, S. and {P{\"u}hlhofer}, G. and {Punch}, M. and {Quirrenbach}, A. and {Raubenheimer}, B.~C. and {Raue}, M. and {Rayner}, S.~M. and {Renaud}, M. and {Rieger}, F. and {Ripken}, J. and {Rob}, L. and {Rosier-Lees}, S. and {Rowell}, G. and {Rudak}, B. and {Rulten}, C.~B. and {Ruppel}, J. and {Sahakian}, V. and {Santangelo}, A. and {Schlickeiser}, R. and {Sch{\"o}ck}, F.~M. and {Schr{\"o}der}, R. and {Schwanke}, U. and {Schwarzburg}, S. and {Schwemmer}, S. and {Shalchi}, A. and {Sikora}, M. and {Skilton}, J.~L. and {Sol}, H. and {Spangler}, D. and {Stawarz}, {\L}. and {Steenkamp}, R. and {Stegmann}, C. and {Superina}, G. and {Szostek}, A. and {Tam}, P.~H. and {Tavernet}, J. -P. and {Terrier}, R. and {Tibolla}, O. and {Tluczykont}, M. and {van Eldik}, C. and {Vasileiadis}, G. and {Venter}, C. and {Venter}, L. and {Vialle}, J.~P. and {Vincent}, P. and {Vink}, J. and {Vivier}, M. and {V{\"o}lk}, H.~J. and {Volpe}, F. and {Wagner}, S.~J. and {Ward}, M. and {Zdziarski}, A.~A. and {Zech}, A.},
        title = "{Discovery of Very High Energy {\ensuremath{\gamma}}-Ray Emission from Centaurus a with H.E.S.S.}",
      journal = {\apjl},
         year = 2009,
        month = apr,
       volume = {695},
       number = {1},
        pages = {L40-L44},
          doi = {10.1088/0004-637X/695/1/L40},
archivePrefix = {arXiv},
       eprint = {0903.1582},
 primaryClass = {astro-ph.CO},
       adsurl = {https://ui.adsabs.harvard.edu/abs/2009ApJ...695L..40A}
}

@ARTICLE{2008ApJ...673...96A,
       author = {{Ajello}, M. and {Rau}, A. and {Greiner}, J. and {Kanbach}, G. and {Salvato}, M. and {Strong}, A.~W. and {Barthelmy}, S.~D. and {Gehrels}, N. and {Markwardt}, C.~B. and {Tueller}, J.},
        title = "{The Swift BAT X-Ray Survey. III. X-Ray Spectra and Statistical Properties}",
      journal = {\apj},
         year = 2008,
        month = jan,
       volume = {673},
       number = {1},
        pages = {96-113},
          doi = {10.1086/524104},
archivePrefix = {arXiv},
       eprint = {0709.4333},
 primaryClass = {astro-ph},
       adsurl = {https://ui.adsabs.harvard.edu/abs/2008ApJ...673...96A}
}

@ARTICLE{1987ApJ...322..643B,
       author = {{Biermann}, P.~L. and {Strittmatter}, P.~A.},
        title = "{Synchrotron Emission from Shock Waves in Active Galactic Nuclei}",
      journal = {\apj},
         year = 1987,
        month = nov,
       volume = {322},
        pages = {643},
          doi = {10.1086/165759},
       adsurl = {https://ui.adsabs.harvard.edu/abs/1987ApJ...322..643B}
}

@ARTICLE{2020NatAs...4..124B,
       author = {{Biteau}, J. and {Prandini}, E. and {Costamante}, L. and {Lemoine}, M. and {Padovani}, P. and {Pueschel}, E. and {Resconi}, E. and {Tavecchio}, F. and {Taylor}, A. and {Zech}, A.},
        title = "{Progress in unveiling extreme particle acceleration in persistent astrophysical jets}",
      journal = {Nature Astronomy},
         year = 2020,
        month = feb,
       volume = {4},
        pages = {124-131},
          doi = {10.1038/s41550-019-0988-4},
archivePrefix = {arXiv},
       eprint = {2001.09222},
 primaryClass = {astro-ph.HE},
       adsurl = {https://ui.adsabs.harvard.edu/abs/2020NatAs...4..124B}
}

@ARTICLE{1984ARA&A..22..319B,
       author = {{Bridle}, Alan H. and {Perley}, Richard A.},
        title = "{Extragalactic Radio Jets}",
      journal = {\araa},
         year = 1984,
        month = jan,
       volume = {22},
        pages = {319-358},
          doi = {10.1146/annurev.aa.22.090184.001535},
       adsurl = {https://ui.adsabs.harvard.edu/abs/1984ARA&A..22..319B}
}

@ARTICLE{2009ApJ...703L.104B,
       author = {{Broderick}, Avery E. and {Loeb}, Abraham},
        title = "{Signatures of Relativistic Helical Motion in the Rotation Measures of Active Galactic Nucleus Jets}",
      journal = {\apjl},
         year = 2009,
        month = oct,
       volume = {703},
       number = {2},
        pages = {L104-L108},
          doi = {10.1088/0004-637X/703/2/L104},
archivePrefix = {arXiv},
       eprint = {0908.2999},
 primaryClass = {astro-ph.HE},
       adsurl = {https://ui.adsabs.harvard.edu/abs/2009ApJ...703L.104B}
}

@ARTICLE{2003MNRAS.345L..40B,
       author = {{Brunetti}, G. and {Mack}, K. -H. and {Prieto}, M.~A. and {Varano}, S.},
        title = "{In-situ particle acceleration in extragalactic radio hot spots: observations meet expectations}",
      journal = {\mnras},
         year = 2003,
        month = nov,
       volume = {345},
       number = {3},
        pages = {L40-L44},
          doi = {10.1046/j.1365-8711.2003.07185.x},
archivePrefix = {arXiv},
       eprint = {astro-ph/0309290},
 primaryClass = {astro-ph},
       adsurl = {https://ui.adsabs.harvard.edu/abs/2003MNRAS.345L..40B}
}

@ARTICLE{1983ApJ...273..128B,
       author = {{Burns}, J.~O. and {Feigelson}, E.~D. and {Schreier}, E.~J.},
        title = "{The inner radio structure of Centaurus A : clues to the origin of thejet X-ray emission.}",
      journal = {\apj},
         year = 1983,
        month = oct,
       volume = {273},
        pages = {128-153},
          doi = {10.1086/161353},
       adsurl = {https://ui.adsabs.harvard.edu/abs/1983ApJ...273..128B}
}

@ARTICLE{1957ApJ...126..457C,
       author = {{Chandrasekhar}, S. and {Kendall}, P.~C.},
        title = "{On Force-Free Magnetic Fields.}",
      journal = {\apj},
         year = 1957,
        month = sep,
       volume = {126},
        pages = {457},
          doi = {10.1086/146413},
       adsurl = {https://ui.adsabs.harvard.edu/abs/1957ApJ...126..457C}
}

@ARTICLE{2001MNRAS.324L..33C,
       author = {{Chiaberge}, M. and {Capetti}, A. and {Celotti}, A.},
        title = "{The BL Lac heart of Centaurus A}",
      journal = {\mnras},
         year = 2001,
        month = jul,
       volume = {324},
       number = {4},
        pages = {L33-L37},
          doi = {10.1046/j.1365-8711.2001.04642.x},
archivePrefix = {arXiv},
       eprint = {astro-ph/0105159},
 primaryClass = {astro-ph},
       adsurl = {https://ui.adsabs.harvard.edu/abs/2001MNRAS.324L..33C}
}

@ARTICLE{2005ApJ...628..104C,
       author = {{Cheung}, C.~C. and {Wardle}, J.~F.~C. and {Chen}, Tingdong},
        title = "{Discovery of Optical Emission in the Hot Spots of Three 3CR Quasars: High-Energy Particle Acceleration in Powerful Radio Hot Spots}",
      journal = {\apj},
         year = 2005,
        month = jul,
       volume = {628},
       number = {1},
        pages = {104-112},
          doi = {10.1086/430634},
archivePrefix = {arXiv},
       eprint = {astro-ph/0503634},
 primaryClass = {astro-ph},
       adsurl = {https://ui.adsabs.harvard.edu/abs/2005ApJ...628..104C}
}

@ARTICLE{1983RPPh...46..973D,
       author = {{Drury}, L. Oc.},
        title = "{REVIEW ARTICLE: An introduction to the theory of diffusive shock acceleration of energetic particles in tenuous plasmas}",
      journal = {Reports on Progress in Physics},
         year = 1983,
        month = aug,
       volume = {46},
       number = {8},
        pages = {973-1027},
          doi = {10.1088/0034-4885/46/8/002},
       adsurl = {https://ui.adsabs.harvard.edu/abs/1983RPPh...46..973D}
}

@ARTICLE{2023NatAs...7.1359F,
       author = {{Fuentes}, Antonio and {G{\'o}mez}, Jos{\'e} L. and {Mart{\'\i}}, Jos{\'e} M. and {Perucho}, Manel and {Zhao}, Guang-Yao and {Lico}, Rocco and {Lobanov}, Andrei P. and {Bruni}, Gabriele and {Kovalev}, Yuri Y. and {Chael}, Andrew and {Akiyama}, Kazunori and {Bouman}, Katherine L. and {Sun}, He and {Cho}, Ilje and {Traianou}, Efthalia and {Toscano}, Teresa and {Dahale}, Rohan and {Foschi}, Marianna and {Gurvits}, Leonid I. and {Jorstad}, Svetlana and {Kim}, Jae-Young and {Marscher}, Alan P. and {Mizuno}, Yosuke and {Ros}, Eduardo and {Savolainen}, Tuomas},
        title = "{Filamentary structures as the origin of blazar jet radio variability}",
      journal = {Nature Astronomy},
         year = 2023,
        month = nov,
       volume = {7},
        pages = {1359-1367},
          doi = {10.1038/s41550-023-02105-7},
archivePrefix = {arXiv},
       eprint = {2311.01861},
 primaryClass = {astro-ph.HE},
       adsurl = {https://ui.adsabs.harvard.edu/abs/2023NatAs...7.1359F}
}

@ARTICLE{2018NatAs...2..472G,
       author = {{Giovannini}, G. and {Savolainen}, T. and {Orienti}, M. and {Nakamura}, M. and {Nagai}, H. and {Kino}, M. and {Giroletti}, M. and {Hada}, K. and {Bruni}, G. and {Kovalev}, Y.~Y. and {Anderson}, J.~M. and {D'Ammando}, F. and {Hodgson}, J. and {Honma}, M. and {Krichbaum}, T.~P. and {Lee}, S. -S. and {Lico}, R. and {Lisakov}, M.~M. and {Lobanov}, A.~P. and {Petrov}, L. and {Sohn}, B.~W. and {Sokolovsky}, K.~V. and {Voitsik}, P.~A. and {Zensus}, J.~A. and {Tingay}, S.},
        title = "{A wide and collimated radio jet in 3C84 on the scale of a few hundred gravitational radii}",
      journal = {Nature Astronomy},
         year = 2018,
        month = apr,
       volume = {2},
        pages = {472-477},
          doi = {10.1038/s41550-018-0431-2},
archivePrefix = {arXiv},
       eprint = {1804.02198},
 primaryClass = {astro-ph.GA},
       adsurl = {https://ui.adsabs.harvard.edu/abs/2018NatAs...2..472G}
}

@ARTICLE{2006MNRAS.368L..15H,
       author = {{Hardcastle}, M.~J. and {Kraft}, R.~P. and {Worrall}, D.~M.},
        title = "{The infrared jet in Centaurus A: multiwavelength constraints on emission mechanisms and particle acceleration}",
      journal = {\mnras},
         year = 2006,
        month = may,
       volume = {368},
       number = {1},
        pages = {L15-L19},
          doi = {10.1111/j.1745-3933.2006.00146.x},
archivePrefix = {arXiv},
       eprint = {astro-ph/0601421},
 primaryClass = {astro-ph},
       adsurl = {https://ui.adsabs.harvard.edu/abs/2006MNRAS.368L..15H}
}

@ARTICLE{2003ApJ...593..169H,
       author = {{Hardcastle}, M.~J. and {Worrall}, D.~M. and {Kraft}, R.~P. and {Forman}, W.~R. and {Jones}, C. and {Murray}, S.~S.},
        title = "{Radio and X-Ray Observations of the Jet in Centaurus A}",
      journal = {\apj},
         year = 2003,
        month = aug,
       volume = {593},
       number = {1},
        pages = {169-183},
          doi = {10.1086/376519},
archivePrefix = {arXiv},
       eprint = {astro-ph/0304443},
 primaryClass = {astro-ph},
       adsurl = {https://ui.adsabs.harvard.edu/abs/2003ApJ...593..169H}
}

@ARTICLE{2010PASA...27..457H,
       author = {{Harris}, Gretchen L.~H. and {Rejkuba}, Marina and {Harris}, William E.},
        title = "{The Distance to NGC 5128 (Centaurus A)}",
      journal = {\pasa},
         year = 2010,
        month = oct,
       volume = {27},
       number = {4},
        pages = {457-462},
          doi = {10.1071/AS09061},
archivePrefix = {arXiv},
       eprint = {0911.3180},
 primaryClass = {astro-ph.GA},
       adsurl = {https://ui.adsabs.harvard.edu/abs/2010PASA...27..457H}
}

@ARTICLE{2000ApJ...545..100H,
       author = {{Hirotani}, Kouichi and {Iguchi}, Satoru and {Kimura}, Moritaka and {Wajima}, Kiyoaki},
        title = "{Pair Plasma Dominance in the Parsec-Scale Relativistic Jet of 3C 345}",
      journal = {\apj},
         year = 2000,
        month = dec,
       volume = {545},
       number = {1},
        pages = {100-106},
          doi = {10.1086/317769},
archivePrefix = {arXiv},
       eprint = {astro-ph/0005394},
 primaryClass = {astro-ph},
       adsurl = {https://ui.adsabs.harvard.edu/abs/2000ApJ...545..100H}
}

@ARTICLE{2004PhRvE..69a6401H,
       author = {{Honda}, M.},
        title = "{Eigenmodes and growth rates of relativistic current filamentation instability in a collisional plasma}",
      journal = {\pre},
         year = 2004,
        month = jan,
       volume = {69},
       number = {1},
          eid = {016401},
        pages = {016401},
          doi = {10.1103/PhysRevE.69.016401},
archivePrefix = {arXiv},
       eprint = {physics/0312144},
 primaryClass = {physics.plasm-ph},
       adsurl = {https://ui.adsabs.harvard.edu/abs/2004PhRvE..69a6401H}
}

@ARTICLE{2008ApJ...675L..61H,
       author = {{Honda}, Mitsuru},
        title = "{Phase-transient Hierarchical Turbulence as an Energy Correlation Generator of Blazar Light Curves}",
      journal = {\apjl},
         year = 2008,
        month = mar,
       volume = {675},
       number = {2},
        pages = {L61},
          doi = {10.1086/533528},
archivePrefix = {arXiv},
       eprint = {0802.0899},
 primaryClass = {astro-ph},
       adsurl = {https://ui.adsabs.harvard.edu/abs/2008ApJ...675L..61H}
}

@ARTICLE{2000Phpl....7.1302H,
       author = {{Honda}, M. and {Meyer-ter-Vehn}, J. and {Pukhov}, A.},
        title = "{Two-dimensional particle-in-cell simulation for magnetized transport of ultra-high relativistic currents in plasma}",
      journal = {Physics of Plasmas},
         year = 2000,
        month = apr,
       volume = {7},
       number = {4},
        pages = {1302-1308},
          doi = {10.1063/1.873941},
       adsurl = {https://ui.adsabs.harvard.edu/abs/2000Phpl....7.1302H}
}

@ARTICLE{2002ApJ...569L..39H,
       author = {{Honda}, Mitsuru and {Honda}, Yasuko S.},
        title = "{Self-Collimation and Magnetic Field Generation of Astrophysical Jets}",
      journal = {\apjl},
         year = 2002,
        month = apr,
       volume = {569},
       number = {1},
        pages = {L39-L42},
          doi = {10.1086/340455},
archivePrefix = {arXiv},
       eprint = {astro-ph/0204048},
 primaryClass = {astro-ph},
       adsurl = {https://ui.adsabs.harvard.edu/abs/2002ApJ...569L..39H}
}

@ARTICLE{2005MNRAS.362..833H,
       author = {{Honda}, Y.~S. and {Honda}, M.},
        title = "{Effects of mirror reflection versus diffusion anisotropy on particle acceleration in oblique shocks}",
      journal = {\mnras},
         year = 2005,
        month = sep,
       volume = {362},
       number = {3},
        pages = {833-837},
          doi = {10.1111/j.1365-2966.2005.09322.x},
archivePrefix = {arXiv},
       eprint = {astro-ph/0510251},
 primaryClass = {astro-ph},
       adsurl = {https://ui.adsabs.harvard.edu/abs/2005MNRAS.362..833H}
}

@ARTICLE{2007ApJ...654..885H,
       author = {{Honda}, Mitsuru and {Honda}, Yasuko S.},
        title = "{Transitive X-Ray Spectrum and PeV Gamma-Ray Cutoff in the M87 Jet: Electron ``Pevatron''}",
      journal = {\apj},
         year = 2007,
        month = jan,
       volume = {654},
       number = {2},
        pages = {885-896},
          doi = {10.1086/509060},
archivePrefix = {arXiv},
       eprint = {astro-ph/0611931},
 primaryClass = {astro-ph},
       adsurl = {https://ui.adsabs.harvard.edu/abs/2007ApJ...654..885H}
}

@ARTICLE{2007ApJ...668..974K,
       author = {{Kato}, Tsunehiko N.},
        title = "{Relativistic Collisionless Shocks in Unmagnetized Electron-Positron Plasmas}",
      journal = {\apj},
         year = 2007,
        month = oct,
       volume = {668},
       number = {2},
        pages = {974-979},
          doi = {10.1086/521297},
archivePrefix = {arXiv},
       eprint = {0707.0545},
 primaryClass = {astro-ph},
       adsurl = {https://ui.adsabs.harvard.edu/abs/2007ApJ...668..974K}
}

@ARTICLE{2002ApJ...569...54K,
       author = {{Kraft}, R.~P. and {Forman}, W.~R. and {Jones}, C. and {Murray}, S.~S. and {Hardcastle}, M.~J. and {Worrall}, D.~M.},
        title = "{Chandra Observations of the X-Ray Jet in Centaurus A}",
      journal = {\apj},
         year = 2002,
        month = apr,
       volume = {569},
       number = {1},
        pages = {54-71},
          doi = {10.1086/339062},
archivePrefix = {arXiv},
       eprint = {astro-ph/0111340},
 primaryClass = {astro-ph},
       adsurl = {https://ui.adsabs.harvard.edu/abs/2002ApJ...569...54K}
}

@ARTICLE{2008A&A...478..111L,
       author = {{Lenain}, J. -P. and {Boisson}, C. and {Sol}, H. and {Katarzy{\'n}ski}, K.},
        title = "{A synchrotron self-Compton scenario for the very high energy {\ensuremath{\gamma}}-ray emission of the radiogalaxy M 87. Unifying the TeV emission of blazars and other AGNs?}",
      journal = {\aap},
         year = 2008,
        month = jan,
       volume = {478},
       number = {1},
        pages = {111-120},
          doi = {10.1051/0004-6361:20077995},
archivePrefix = {arXiv},
       eprint = {0710.2847},
 primaryClass = {astro-ph},
       adsurl = {https://ui.adsabs.harvard.edu/abs/2008A&A...478..111L}
}

@ARTICLE{2017ApJ...842...39L,
       author = {{Liu}, Ruo-Yu and {Rieger}, F.~M. and {Aharonian}, F.~A.},
        title = "{Particle Acceleration in Mildly Relativistic Shearing Flows: The Interplay of Systematic and Stochastic Effects, and the Origin of the Extended High-energy Emission in AGN Jets}",
      journal = {\apj},
         year = 2017,
        month = jun,
       volume = {842},
       number = {1},
          eid = {39},
        pages = {39},
          doi = {10.3847/1538-4357/aa7410},
archivePrefix = {arXiv},
       eprint = {1706.01054},
 primaryClass = {astro-ph.HE},
       adsurl = {https://ui.adsabs.harvard.edu/abs/2017ApJ...842...39L}
}

@BOOK{longair94,
       author = {{Longair}, M.~S.},
        title = "{High Energy Astrophysics: Stars, the Galaxy and the interstellar mdeium}",
    publisher = {Cambridge University Press},
	 year = "1994",
      edition = {2},
	 isbn = {0521434394}
}

@ARTICLE{2000ApJ...528..276M,
       author = {{Marconi}, Alessandro and {Schreier}, Ethan J. and {Koekemoer}, Anton and {Capetti}, Alessandro and {Axon}, David and {Macchetto}, Duccio and {Caon}, Nicola},
        title = "{Unveiling the Active Nucleus of Centaurus A}",
      journal = {\apj},
         year = 2000,
        month = jan,
       volume = {528},
       number = {1},
        pages = {276-291},
          doi = {10.1086/308168},
archivePrefix = {arXiv},
       eprint = {astro-ph/9907378},
 primaryClass = {astro-ph},
       adsurl = {https://ui.adsabs.harvard.edu/abs/2000ApJ...528..276M}
}

@ARTICLE{2007ApJ...665..209M,
       author = {{Markowitz}, A. and {Takahashi}, T. and {Watanabe}, S. and {Nakazawa}, K. and {Fukazawa}, Y. and {Kokubun}, M. and {Makishima}, K. and {Awaki}, H. and {Bamba}, A. and {Isobe}, N. and {Kataoka}, J. and {Madejski}, G. and {Mushotzky}, R. and {Okajima}, T. and {Ptak}, A. and {Reeves}, J.~N. and {Ueda}, Y. and {Yamasaki}, T. and {Yaqoob}, T.},
        title = "{The Suzaku Observation of the Nucleus of the Radio-loud Active Galaxy Centaurus A: Constraints on Abundances of the Accreting Material}",
      journal = {\apj},
         year = 2007,
        month = aug,
       volume = {665},
       number = {1},
        pages = {209-224},
          doi = {10.1086/519271},
archivePrefix = {arXiv},
       eprint = {0704.3743},
 primaryClass = {astro-ph},
       adsurl = {https://ui.adsabs.harvard.edu/abs/2007ApJ...665..209M}
}

@ARTICLE{2025PhRvD.111b3024M,
       author = {{Mbarek}, Rostom and {Caprioli}, Damiano and {Murase}, Kohta},
        title = "{Ultrahigh-energy neutrinos as a probe of espresso-shear acceleration in jets of Centaurus A}",
      journal = {\prd},
         year = 2025,
        month = jan,
       volume = {111},
       number = {2},
          eid = {023024},
        pages = {023024},
          doi = {10.1103/PhysRevD.111.023024},
archivePrefix = {arXiv},
       eprint = {2410.05696},
 primaryClass = {astro-ph.HE},
       adsurl = {https://ui.adsabs.harvard.edu/abs/2025PhRvD.111b3024M}
}

@ARTICLE{1997A&A...325...57M,
       author = {{Meisenheimer}, K. and {Yates}, M.~G. and {Roeser}, H. -J.},
        title = "{The synchrotron spectra of radio hot spots. II. Infrared imaging.}",
      journal = {\aap},
         year = 1997,
        month = sep,
       volume = {325},
        pages = {57-73},
       adsurl = {https://ui.adsabs.harvard.edu/abs/1997A&A...325...57M}
}

@ARTICLE{2007A&A...471..453M,
       author = {{Meisenheimer}, K. and {Tristram}, K.~R.~W. and {Jaffe}, W. and {Israel}, F. and {Neumayer}, N. and {Raban}, D. and {R{\"o}ttgering}, H. and {Cotton}, W.~D. and {Graser}, U. and {Henning}, Th. and {Leinert}, Ch. and {Lopez}, B. and {Perrin}, G. and {Prieto}, A.},
        title = "{Resolving the innermost parsec of Centaurus A at mid-infrared wavelengths}",
      journal = {\aap},
         year = 2007,
        month = aug,
       volume = {471},
       number = {2},
        pages = {453-465},
          doi = {10.1051/0004-6361:20066967},
archivePrefix = {arXiv},
       eprint = {0707.0177},
 primaryClass = {astro-ph},
       adsurl = {https://ui.adsabs.harvard.edu/abs/2007A&A...471..453M}
}

@ARTICLE{1979PhFl...22..866M,
       author = {{Montgomery}, D. and {Liu}, C.~S.},
        title = "{Nonlinear development of an electromagnetic filamentation instability}",
      journal = {Physics of Fluids},
         year = 1979,
        month = may,
       volume = {22},
       number = {5},
        pages = {866-870},
          doi = {10.1063/1.862674},
       adsurl = {https://ui.adsabs.harvard.edu/abs/1979PhFl...22..866M}
}

@ARTICLE{2014A&A...562A..12P,
       author = {{Petropoulou}, M. and {Lefa}, E. and {Dimitrakoudis}, S. and {Mastichiadis}, A.},
        title = "{One-zone synchrotron self-Compton model for the core emission of Centaurus A revisited}",
      journal = {\aap},
         year = 2014,
        month = feb,
       volume = {562},
          eid = {A12},
        pages = {A12},
          doi = {10.1051/0004-6361/201322833},
archivePrefix = {arXiv},
       eprint = {1311.1119},
 primaryClass = {astro-ph.HE},
       adsurl = {https://ui.adsabs.harvard.edu/abs/2014A&A...562A..12P}
}

@ARTICLE{2024A&A...692A..48R,
       author = {{Raiteri}, C.~M. and {Villata}, M. and {Carnerero}, M.~I. and {Kurtanidze}, S.~O. and {Mirzaqulov}, D.~O. and {Ben{\'\i}tez}, E. and {Bonnoli}, G. and {Carosati}, D. and {Acosta-Pulido}, J.~A. and {Agudo}, I. and {Andreeva}, T.~S. and {Apolonio}, G. and {Bachev}, R. and {Borman}, G.~A. and {Bozhilov}, V. and {Brown}, L.~F. and {Carbonell}, W. and {Casadio}, C. and {Chen}, W.~P. and {Damljanovic}, G. and {Ehgamberdiev}, S.~A. and {Elsaesser}, D. and {Escudero}, J. and {Feige}, M. and {Fuentes}, A. and {Gabellini}, D. and {Gazeas}, K. and {Giroletti}, M. and {Grishina}, T.~S. and {Gupta}, A.~C. and {Gurwell}, M.~A. and {Hagen-Thorn}, V.~A. and {Hamed}, G.~M. and {Hiriart}, D. and {Hodges}, M. and {Ivanidze}, R.~Z. and {Ivanov}, D.~V. and {Joner}, M.~D. and {Jorstad}, S.~G. and {Jovanovic}, M.~D. and {Kiehlmann}, S. and {Kimeridze}, G.~N. and {Kopatskaya}, E.~N. and {Kovalev}, Yu. A. and {Kovalev}, Y.~Y. and {Kurtanidze}, O.~M. and {Kurtenkov}, A. and {Larionova}, E.~G. and {Lessing}, A. and {Lin}, H.~C. and {L{\'o}pez}, J.~M. and {Lorey}, C. and {Ludwig}, J. and {Marchili}, N. and {Marchini}, A. and {Marscher}, A.~P. and {Matsumoto}, K. and {Max-Moerbeck}, W. and {Mihov}, B. and {Minev}, M. and {Mingaliev}, M.~G. and {Modaressi}, A. and {Morozova}, D.~A. and {Mortari}, F. and {Mufakharov}, T.~V. and {Myserlis}, I. and {Nikolashvili}, M.~G. and {Pearson}, T.~J. and {Popkov}, A.~V. and {Rahimov}, I.~A. and {Readhead}, A.~C.~S. and {Reinhart}, D. and {Reeves}, R. and {Righini}, S. and {Romanov}, F.~D. and {Savchenko}, S.~S. and {Semkov}, E. and {Shishkina}, E.~V. and {Sigua}, L.~A. and {Slavcheva-Mihova}, L. and {Sotnikova}, Yu. V. and {Steineke}, R. and {Stojanovic}, M. and {Strigachev}, A. and {Takey}, A. and {Traianou}, E. and {Troitskaya}, Yu. V. and {Troitskiy}, I.~S. and {Tsai}, A.~L. and {Valcheva}, A. and {Vasilyev}, A.~A. and {Verna}, G. and {Vince}, O. and {Vrontaki}, K. and {Weaver}, Z.~R. and {Webb}, J. and {Yuldoshev}, Q.~X. and {Zaharieva}, E. and {Zhovtan}, A.~V.},
        title = "{A wiggling filamentary jet at the origin of the blazar multi-wavelength behaviour}",
      journal = {\aap},
         year = 2024,
        month = dec,
       volume = {692},
          eid = {A48},
        pages = {A48},
          doi = {10.1051/0004-6361/202452311},
archivePrefix = {arXiv},
       eprint = {2410.22319},
 primaryClass = {astro-ph.HE},
       adsurl = {https://ui.adsabs.harvard.edu/abs/2024A&A...692A..48R}
}

@ARTICLE{2011ApJ...733...23R,
       author = {{Rothschild}, R.~E. and {Markowitz}, A. and {Rivers}, E. and {Suchy}, S. and {Pottschmidt}, K. and {Kadler}, M. and {M{\"u}ller}, C. and {Wilms}, J.},
        title = "{Twelve and a Half Years of Observations of Centaurus a with the Rossi X-Ray Timing Explorer}",
      journal = {\apj},
         year = 2011,
        month = may,
       volume = {733},
       number = {1},
          eid = {23},
        pages = {23},
          doi = {10.1088/0004-637X/733/1/23},
archivePrefix = {arXiv},
       eprint = {1102.5076},
 primaryClass = {astro-ph.HE},
       adsurl = {https://ui.adsabs.harvard.edu/abs/2011ApJ...733...23R}
}

@BOOK{Schlickeiser02,
       author = {{Schlickeiser}, R.},
        title = "{Cosmic Ray Astrophysics}",
    publisher = {Springer: Berlin},
	 year = "2002",
      edition = {1},
	 isbn = {3540664653}
}

@ARTICLE{2003ApJ...596L.121S,
       author = {{Silva}, L.~O. and {Fonseca}, R.~A. and {Tonge}, J.~W. and {Dawson}, J.~M. and {Mori}, W.~B. and {Medvedev}, M.~V.},
        title = "{Interpenetrating Plasma Shells: Near-equipartition Magnetic Field Generation and Nonthermal Particle Acceleration}",
      journal = {\apjl},
         year = 2003,
        month = oct,
       volume = {596},
       number = {1},
        pages = {L121-L124},
          doi = {10.1086/379156},
archivePrefix = {arXiv},
       eprint = {astro-ph/0307500},
 primaryClass = {astro-ph},
       adsurl = {https://ui.adsabs.harvard.edu/abs/2003ApJ...596L.121S}
}

@ARTICLE{2019ApJ...871..248S,
       author = {{Snios}, Bradford and {Wykes}, Sarka and {Nulsen}, Paul E.~J. and {Kraft}, Ralph P. and {Meyer}, Eileen T. and {Birkinshaw}, Mark and {Worrall}, Diana M. and {Hardcastle}, Martin J. and {Roediger}, Elke and {Forman}, William R. and {Jones}, Christine},
        title = "{Variability and Proper Motion of X-Ray Knots in the Jet of Centaurus A}",
      journal = {\apj},
         year = 2019,
        month = feb,
       volume = {871},
       number = {2},
          eid = {248},
        pages = {248},
          doi = {10.3847/1538-4357/aafaf3},
archivePrefix = {arXiv},
       eprint = {1901.00509},
 primaryClass = {astro-ph.HE},
       adsurl = {https://ui.adsabs.harvard.edu/abs/2019ApJ...871..248S}
}

@ARTICLE{1998A&A...330...97S,
       author = {{Steinle}, H. and {Bennett}, K. and {Bloemen}, H. and {Collmar}, W. and {Diehl}, R. and {Hermsen}, W. and {Lichti}, G.~G. and {Morris}, D. and {Schonfelder}, V. and {Strong}, A.~W. and {Williams}, O.~R.},
        title = "{COMPTEL observations of Centaurus A at MeV energies in the years 1991 to 1995}",
      journal = {\aap},
         year = 1998,
        month = feb,
       volume = {330},
        pages = {97-107},
       adsurl = {https://ui.adsabs.harvard.edu/abs/1998A&A...330...97S}
}

@ARTICLE{2016PASJ...68S..25T,
       author = {{Tachibana}, Yutaro and {Kawamuro}, Taiki and {Ueda}, Yoshihiro and {Shidatsu}, Megumi and {Arimoto}, Makoto and {Yoshii}, Taketoshi and {Yatsu}, Yoichi and {Saito}, Yoshihiko and {Pike}, Sean and {Kawai}, Nobuyuki},
        title = "{A soft X-ray lag detected in Centaurus A}",
      journal = {\pasj},
         year = 2016,
        month = jun,
       volume = {68},
          eid = {S25},
        pages = {S25},
          doi = {10.1093/pasj/psw001},
archivePrefix = {arXiv},
       eprint = {1504.03208},
 primaryClass = {astro-ph.HE},
       adsurl = {https://ui.adsabs.harvard.edu/abs/2016PASJ...68S..25T}
}

@ARTICLE{2019ApJ...878..139T,
       author = {{Tanada}, K. and {Kataoka}, J. and {Inoue}, Y.},
        title = "{Inverse Compton Scattering of Starlight in the Kiloparsec-scale Jet in Centaurus A: The Origin of Excess TeV {\ensuremath{\gamma}}-Ray Emission}",
      journal = {\apj},
         year = 2019,
        month = jun,
       volume = {878},
       number = {2},
          eid = {139},
        pages = {139},
          doi = {10.3847/1538-4357/ab2233},
archivePrefix = {arXiv},
       eprint = {1905.07055},
 primaryClass = {astro-ph.HE},
       adsurl = {https://ui.adsabs.harvard.edu/abs/2019ApJ...878..139T}
}
\end{document}